\documentclass[iop]{emulateapj}

\usepackage{apjfonts}
\usepackage{pbox}
\usepackage{bm}
\usepackage{CJK}
\usepackage{color}
\usepackage{amsmath}

\newcommand{\DJz}{\ensuremath{\widehat{\Delta J_z}_{\, {\rm 1Gyr}}}}

\newcommand{\ageGyr}{\Bigl (\frac{\tau}{1{\rm Gyr}}\Bigr )} 
\newcommand{\Rbar}{\ensuremath{\overline{R}_{\rm GC}}}
\newcommand{\gammRC}{\ensuremath{\gamma(\overline{R}_{\rm GC})}}

\shorttitle{vertical motion history of disk stars}
\shortauthors{Ting \& Rix}

\begin{document}

\begin{CJK*}{UTF8}{gbsn}
\title{The vertical motion history of disk stars throughout the Galaxy}
\author{Yuan-Sen Ting (丁源森)\altaffilmark{1,2,3,4,5} \& Hans-Walter Rix\altaffilmark{4}}
\altaffiltext{1}{Institute for Advanced Study, Princeton, NJ 08540, USA}
\altaffiltext{2}{Department of Astrophysical Sciences, Princeton University, Princeton, NJ 08544, USA}
\altaffiltext{3}{Observatories of the Carnegie Institution of Washington, 813 Santa Barbara Street, Pasadena, CA 91101, USA}
\altaffiltext{4}{Max Planck Institute for Astronomy, K\"onigstuhl 17, D-69117 Heidelberg, Germany}
\altaffiltext{5}{Hubble Fellow}
\slugcomment{ApJ, accepted for publication -- 2019 April 29}

%
%
%
%
%
%

\begin{abstract}
It has long been known that the vertical motions of Galactic disk stars increase with stellar age, commonly interpreted as vertical heating through orbit scattering. Here we map the vertical actions of disk stars as a function of age ($\tau\le 8$~Gyrs) and across a large range of Galactocentric radii, $\overline{R}_{\rm GC}$, drawing on APOGEE and Gaia data. We fit $\widehat{J_z}(\overline{R}_{\rm GC},\tau )$ as a combination of the vertical action at birth, $\widehat{J_{z,0}}$, and subsequent heating $\DJz(\overline{R}_{\rm GC})$ that scales as $\tau^{\gamma(\overline{R}_{\rm GC})}$. The inferred birth temperature, $\widehat{J_{z,0}}(\overline{R}_{\rm GC})$ is $1 \;{\rm kpc}\;{\rm km/s}$ for $3~{\rm kpc}< \overline{R}_{\rm GC}<10~{\rm kpc}$, consistent with the ISM velocity dispersion; but it rapidly rises outward, to $8\;{\rm kpc}\;{\rm km/s}$ for $\overline{R}_{\rm GC} = 14~{\rm kpc}$, likely reflecting the stars' birth in a warped or flared gas disk. We find the heating rate $\DJz$ to be modest and nearly constant across all radii, $1.6\;{\rm kpc}\;{\rm km/s}\;{\rm Gyr}^{-1}$. The stellar age dependence $\gamma$ gently grows with Galactocentric radius, from $\gamma \simeq 1$ for $\overline{R}_{\rm GC}\lesssim R_\odot$ to $\gamma \simeq 1.3$ at $\overline{R}_{\rm GC} =14\,$kpc. The observed $J_z - \tau$ relation at all radii is considerably steeper ($\gamma\gtrsim 1$) than the time dependence theoretically expected from orbit scattering, $J_z\propto t^{0.5}$. We illustrate how this conundrum can be resolved if we also account for the fact that at earlier epochs the scatterers were more common, and the restoring force from the stellar disk surface mass density was low. Our analysis may re-instate gradual orbital scattering as a plausible and viable mechanism to explain the age-dependent vertical motions of disk stars. 
\end{abstract}


%
%
%
%
%
%

\section{Introduction}
\label{sec:introduction}

In the context of $\Lambda$CDM hierarchical cosmogony, galaxy formation started out vigorously, with a rapid gas inflow and frequent mergers \citep[e.g.,][]{whi78,bro04,bro12,vog14}. During this period, stars formed from highly turbulent, clumpy ISM with large velocity dispersion and were born with kinematically hot orbits \citep{bou09,for12}, as also borne out by high redshift observations \citep[e.g.,][]{for09,win15}. For the Milky Way, this early phase of vigorous evolution faded 8--10 Gyrs ago, giving way to more gradual gas acquisition and an extended period without any major mergers. Since then, gas increasingly settled into a thin disk before forming stars, resulting in ``upside-down'' formation of the main, extended\footnote{This component is loosely referred to as the low-$\alpha$ or ``thin'' disk; it is this disk ``component'' that is the subject of this work.} stellar disk \citep{bir13,sti13,gra16}. At present, the star-forming molecular gas in the Milky Way disk has a small velocity dispersion \citep{sta89,sta05,sta06,aum09,mar14,aum16}, and so do the stars that form from it. Subsequently, stars are bound to acquire more vertical motion over time through a variety of dynamical processes. Indeed, vertical thickening of spiral disks seems prevalent throughout the cosmos \citep[e.g.,][]{yoa06,jur08}. The present day distribution of vertical motions, e.g., characterized by their velocity dispersion, $\sigma_z$ in stars of different ages and Galactocentric radii therefore reflects a combination of their birth ``temperature'' and subsequent heating. Constraining and understanding the vertical motions of Galactic disk stars therefore provides a key test of the processes presumed to set the vertical structure of disks in general. 

Several different physical processes may have contributed to the vertical heating of the Galactic stellar disk, causing either rapid ``non-adiabatic heating'', or more gradual ``adiabatic heating''. Cosmological simulations have shown that galaxies of the Milky Way's size are frequently experiencing minor mergers \citep{qui93,wal96,vel99,kaz09,hou11,gom13,don16,moe16}, external perturbations that can heat up the disk. But there are also more gradual internal heating mechanism: classically, giant molecular clouds (GMCs) act as scatterers that could heat up the disk either directly or by redirecting some of the in-plane heating (e.g., through transient spiral arms or the Galactic bar) to the vertical direction \citep{spi53,bar67,lac84,car85,car87,jen90}. In-plane random motions can also be converted to vertical motions in the stellar disk through bending waves \citep[e.g.,][]{sha10,fau14,deb14,wid14}.

But the interpretation of disk stars' vertical velocity dispersion, $\sigma_z$ as disk heating is often complicated by the overall secular evolution of the disk. For instance, the gradual, adiabatic increase in the mid-plane baryon density will cause an increase in $\sigma_z$ \citep[e.g.,][]{bah84,van88,jen92,vil10}. And radial migration of stars \citep{sel02,sch09b,min10} also affects the vertical disk structure: it should cause more extended vertical motion at reduced velocity dispersion for stars that move outward, and the opposite effect for stars that move inward \citep{loe11,min12,ros13,mar14,ver14,aum17}. 

Traditionally, studies have characterized the effects of vertical disk heating through the age-velocity dispersion relation (AVR) \citep[e.g.,][]{str46,wie77,qui01,sea07,sou08,aum09,san18}. The Geneva-Copenhagen Survey (GCS) provides the current state-of-the-art for such a relation in the solar neighborhood \citep{nor04,hol07,hol09,aum09,cas11}. Solar neighborhood studies basically agree that the present-day vertical velocity dispersion among stars of age $\tau$ scales approximately as $\sigma_z \sim \tau^{0.5}$. But the interpretation of this scaling is still under debate, as the simplest models of heating through a time-independent population of scatterers would imply a different evolution of vertical motions $\sigma_z \sim \tau^{0.25}$ \citep{lac84,han02}. And with only local information at hand, the various heating mechanisms, when combined with the effects of radial migration, may have degenerate observational signatures. 

The combination of Gaia DR2 \citep{gai18,lin18}, of the on-going large-scale spectroscopic surveys, such as APOGEE \citep{maj17}, Galah \citep{des15} and Gaia-ESO \citep{smi14}, and of consistent stellar age estimates across a wide range of Galactocentric radii \citep[e.g.,][]{mar15,nes16} make it now possible to characterize and interpret the history of vertical motions among (low-$\alpha$) disk stars throughout the Galaxy. This is what we set out to do in this paper.

It seems sensible to characterize the vertical motions of stars by their vertical actions, $J_z$, rather than their vertical velocities $v_z$ or velocity dispersion, $\sigma_z$. This is mainly because the vertical action is an adiabatic invariant under any gradual changes of the potential reflecting the growth of the Galaxy, and an approximate invariant under radial migration through churning \citep{car87,sel13}, as the vertical motions are only weakly coupled to in-plane resonances. In contrast, $\sigma_z$ will change in a growing potential and under radial migration, with the scale-height h$_z$ changing in compensation such as to conserve $J_z$. Note that for the simple case of harmonic vertical oscillations, $\sigma_z\propto \tau^{0.25}$ corresponds to $J_z\propto \tau^{0.5}$. Therefore, mapping the present-day $J_z$ as a function of stellar age $\tau$ and current Galactocentric radius, may make it easier to interpret this distribution in terms of: (a) with what ``vertical birth temperature'' did star start out, and (b) what subsequent heating (defined as an increase in $J_z$) did they incur subsequently.

Specifically, we combine here a red clump sample derived from APOGEE in \citet{tin18a} with their Gaia DR2 proper motions \citep{lin18}, which yield a large number of stars across the Milky Way disk with precise 6D phase space coordinates, from which their vertical actions can be reliably estimated. Furthermore, precise stellar ages ($25\%$) can be derived for this sample as we will show. A global ($3~{\rm kpc}<\overline{R}_{\rm GC}<14~{\rm kpc}$) study of the vertical structure of the Milky Way main stellar disk\footnote{In this study, we use the term main, extended or thin stellar disk to describe the low-$\alpha$ sequence of the Galactic disk \citep[e.g.,][]{rix13,hay15}, and will use these terms interchangeably. Similarly, the thick disk in this study describes the $\alpha$-enhanced disk and does not necessarily mean the geometrical thick disk.} also requires careful modeling of the selection function. 

The paper is organized as follows. In Section~\ref{sec:data}, we present and derive precise distances and actions for the red clump sample, as well as precise ages, inferred from a data-driven model drawing asteroseismic training data. In Section~\ref{sec:model}, we introduce a physical model of vertical action which capture the full distribution of vertical action at different radii and ages. We also describe the derivation of the selection function in vertical action, which turns out to be important in the subsequent Bayesian inference. In Section~\ref{sec:results}, we will discuss the inference posterior of our models and study how the vertical heating and birth temperature of stars depends on the Galactocentric radii. We will discuss the results in the context of a simple epoch-dependent but analytic scattering model in Section~\ref{sec:discussion} and conclude in Section~\ref{sec:conclusion}. Throughout this paper, we will assume the following units: velocity [km/s], distance [kpc], age/time [Gyr], and vertical action [kpc km/s].


%
%
%
%
%

\section{Data}
\label{sec:data}

In this study, we adopt the APOGEE red clump sample derived in \citet{tin18a}, and cross-match the sample with Gaia DR2 \citep{gai18,lin18} astrometry to obtain proper motions. We do not use the Gaia DR2 data for distance determinations, as for stars beyond a few kpc the Gaia 1/parallax ($1/\varpi$) distance precision degrades drastically and becomes increasingly biased \citep{bai18}. Fortunately, red clump stars are excellent standard candles regardless of their metallicities and stellar ages, and mitigate this problem; across the Milky Way disk, they can yield photometric distances precise to about $\sigma_d/d = 7\%$ \citep{bov14,haw17}, where $d$ is the heliocentric distance.

The red clump sample was selected among APOGEE DR14 targets through a deep learning algorithm which optimizes an empirical data-driven mapping from the APOGEE normalized spectra to their corresponding asteroseismic period spacing $\Delta P$ and frequency separation $\Delta \nu$, adopting the \citet{vra16} asteroseismic data as the training set \citep{tin18a}. \citet{haw18} and \citet{tin18a} showed that the inferred mixed mode period separation $\Delta P$ can yield a more pristine red clump sample with only $2-3\%$ contamination for APOGEE, at least two times lower in contamination than previous approaches \citep[e.g.,][]{bov14}; see \citet{tin18a} for more details.

\begin{figure}
\centering
\includegraphics[width=0.5\textwidth]{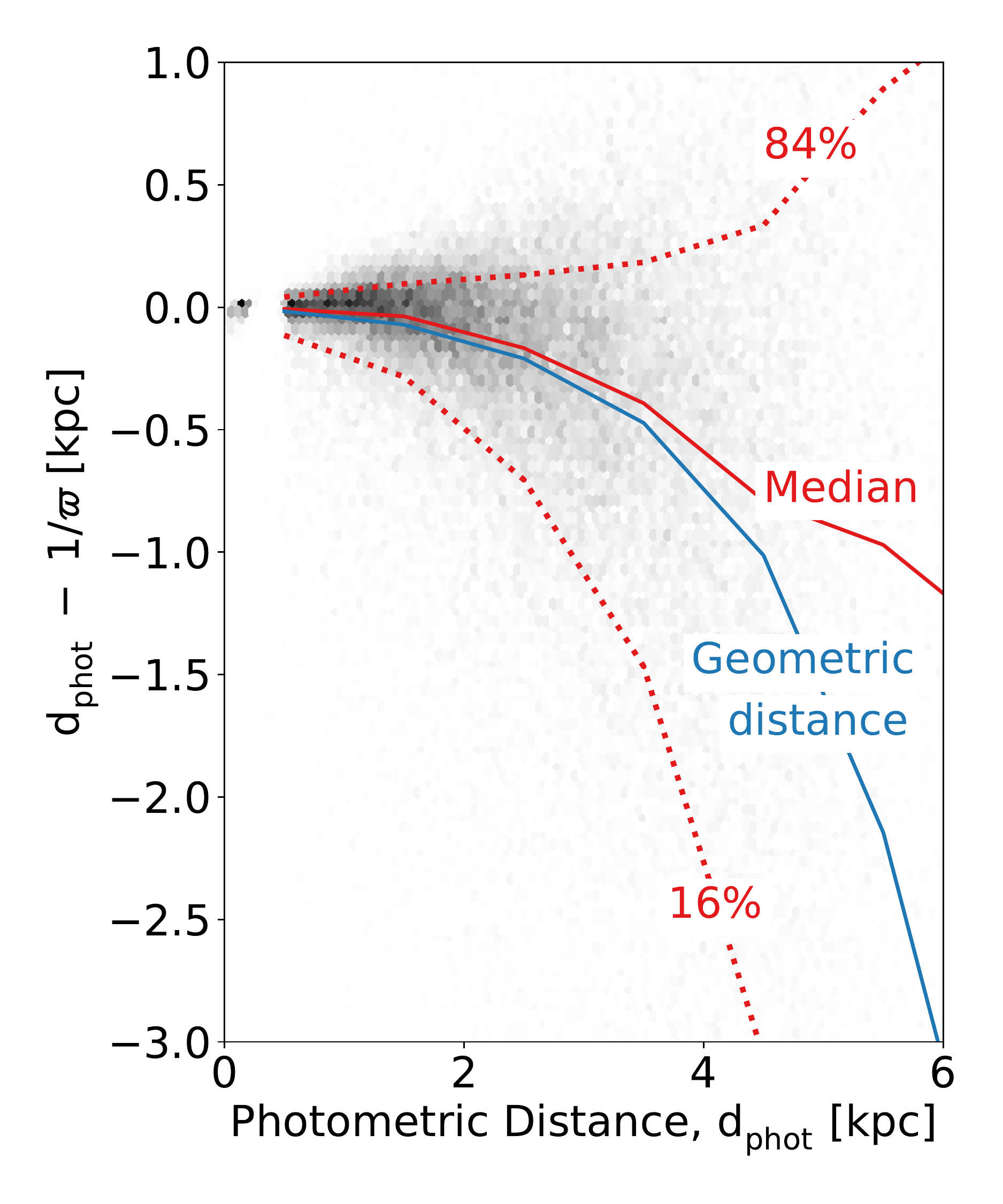}
\caption{Red clump stars in this study have precise photometric distances. The Figure shows the comparison between the derived photometric distances $d_{\rm phot}$ to the Gaia $1/\varpi$ parallax distance. As shown in \citet{bai18} (blue line) which is based on the Milky Way geometric prior, as we move further away from the Sun, $1/\varpi$ yields poor and biased distances. The photometric distances agree with the geometric distances but are likely to be more precise ($\sigma_d/d \sim 7\%$) because red clumps are standard candles.}
\label{fig1}
\end{figure}

To estimate photometric distances, we cross-match this red clump sample with the allWISE catalog to get the WISE $W1$ magnitude. We obtain the (small) $W1$ extinction correction from the $G-W1$ color, assuming $A_G/A_{W1} = 16$ \citep{haw17}. From the $G-W1$ color, we also select stars that are less extincted and adopt this subset to fit for the absolute magnitude $M_{W1}$, as a quadratic function of the APOGEE's $T_{\rm eff}$. The variance around this fit shows a spread of $15\%$, corresponding $\sigma_d/d \simeq 7\%$ \citep[see also][]{bov14,haw17}. For each star, we estimate its photometric distance through the predicted $M_{W1}(T_{\rm eff})$ from the fitted relation and correct for the extinction estimated from its $G-W1$ color. Fig.~\ref{fig1} shows the comparison between our photometric distance estimates to Gaia's $1/\varpi$ distances. The Figure illustrates that 1/$\varpi$ is indeed a biased distance indicator at large heliocentric distance, as expected from \citet{bai18}, which is overplotted in blue in the Figure.

\begin{figure}
\centering
\includegraphics[width=0.5\textwidth]{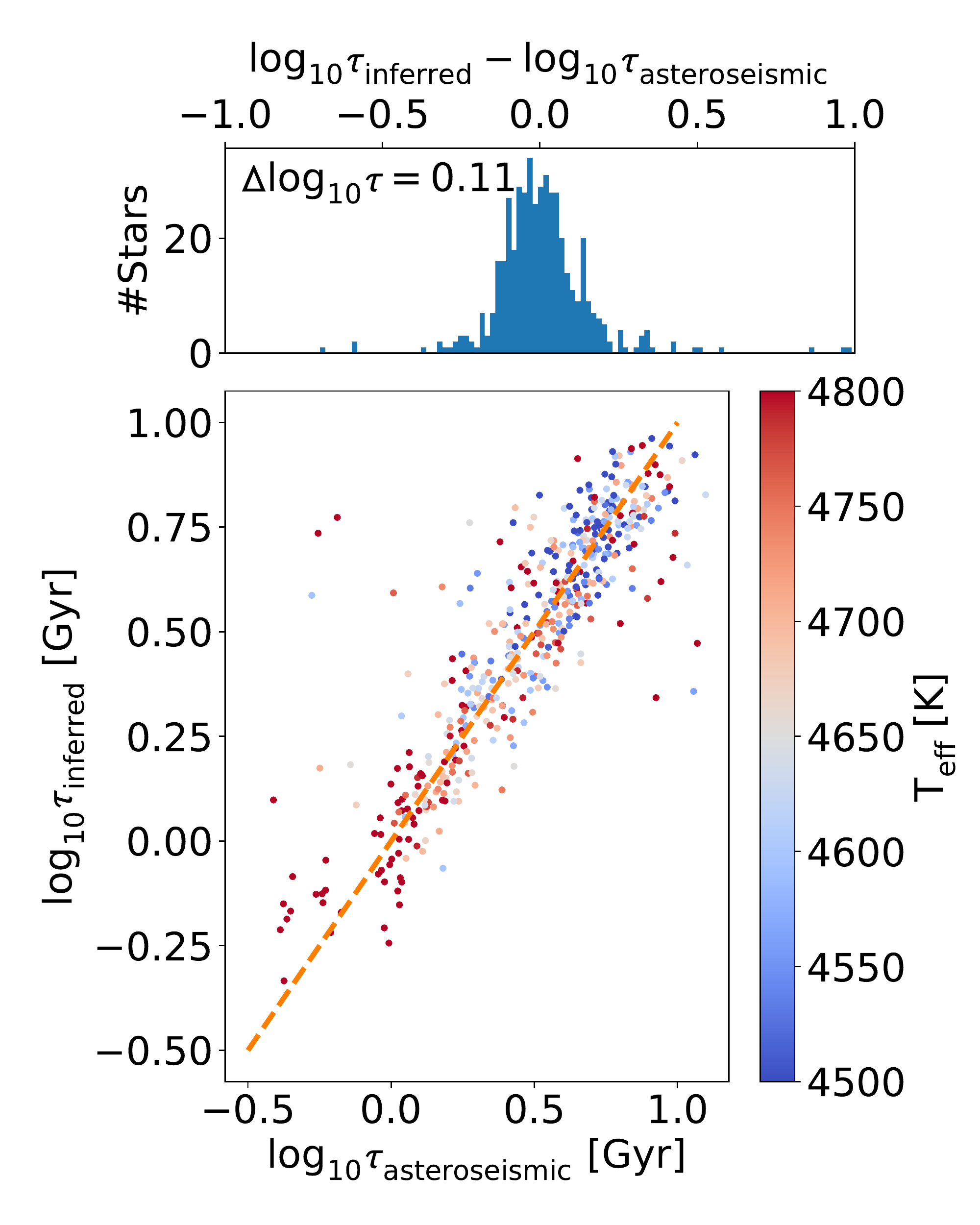}
\caption{Deriving precise spectroscopic ages for red clump stars in this study. We infer stellar spectroscopic ages through APOGEE spectra using APOKASC-2 asteroseismic ages as the training set. The bottom panel shows the cross-validation between the asteroseismic stellar ages derived in APOKASC-2 and their corresponding inferred spectroscopic ages of the test set (not used in training), color-coded with the effective temperature of the stars. The top panel shows the histogram of the deviations. The cross-validation indicates that we have unbiased (with respect to the asteroseismic values) spectroscopic ages across most stellar ages ($\tau = 1-8$ Gyr), independent of the effective temperature of stars, and the spectroscopic ages are precise to $\sim 25\%$ (or $0.11\,$dex).}
\label{fig2}
\end{figure}

To estimate stellar ages, we build a spectroscopic age estimator that is trained on a set of $\sim 2400$ APOKASC-2 determined red clump stars with precise asteroseismic age estimates \citep{pin18} that are also in the \citet{tin18a} catalog. We only select stars that have asteroseismic ages $< 13\,$Gyr. It has been well established that spectra from giants contain information about stellar ages through their mass-dependent dredge up, which affects the photospheric C/N ratios \citep{mar15,nes16}. Rather than making reference to C/N directly, we apply neural networks to optimize such the mapping from spectra to asteroseismic ages \citep[see][]{tin18a}, and then propagate asteroseismic ages to our red clump sample. While we infer stellar ages $\tau$ here, instead of asteroseismic $\Delta P$ and $\Delta \nu$ as in \citet{tin18a}, the algorithm is exactly the same: we establish an empirical, high-dimensional, mapping from the APOGEE normalized spectra to stellar ages via a fully connected neural network with three hidden layers and 50 neurons each; the neurons are connected with a Sigmoid activation function, and the model is optimized using {\sc pytorch}. We cross-validated our inference using a random $400$ stars with asteroseismic ages, comparing them to our model predictions trained on the other $2000$ APOKASC stars, as illustrated in Fig.~\ref{fig2}. The cross-validation indicates that our inferred spectroscopic ages are largely unbiased, at least from $\tau = 1-8\,$Gyr, and precise to $\sigma_\tau/\tau= 25\%$ ($\Delta \log_{10} J_z = 0.11\,$dex); here we assume the half of the 16-84 percentile intervals as $1\sigma$. 

Nonetheless, we caution that, at least with the training set at hand, this empirical mapping method which presumably exploits the C/N information in spectra, does not seem to have enough spectral information to distinguish stars older than $8\,$Gyr and tend to be biased in the oldest regime; stars older than $8\,$Gyr are often incorrectly inferred as $\sim 8\,$Gyr (as can be seen at the oldest age regime in the bottom panel of Fig.~\ref{fig2}). A similar problem also happens for the few stars younger than $1\,$Gyr. But since our study only focuses on stars that are younger than $8\,$Gyr, and the red clump sample is strongly biased against stars older than $8\,$Gyr, we find that this bias only minimally affects our inference.  But we caution audiences who might want to adopt the published catalog in this study as there is a lack of stars older than $8\,$Gyr due to this bias. Similarly, at the young end, we found that only about $3\%$ of our sample are in the very young regime ($\log \tau/{\rm Gyr}$ < -0.2) where the inferred stellar ages might be biased. Further, a small bias in log scale at the lower end, i.e., moving stars from 0.4 Gyr to 0.6 Gyr (see Fig.~\ref{fig2}), should have a negligible impact on our inferences because the heating rate is dominated by the more extended evolutionary behaviors of the stars ($\mathcal{O}(1\;{\rm Gyr}))$.

In this analysis we are interested in the disk heating mechanisms that were effective after the initial vigorous and turbulent formation phase -- extending to a redshift of $z\sim 1$, about 8~Gyrs ago -- when stars were presumably born kinematically hot \citep{nog98,bou09,bro04,bro06,bir13,gra16}. Most $\alpha$-enhanced stars in the disk probably formed in that period. In the subsequent analysis, we limit ourselves to stars with inferred spectroscopic ages $\tau < 8\,$Gyr and in the low-$\alpha$ sequence of the Milky Way disk, with ${\rm [Mg/Fe]}_{\rm Payne} < -0.1\,{\rm [Fe/H]}_{\rm Payne} + 0.18$, where we adopt the APOGEE abundances from a reanalysis of the APOGEE spectra using the spectral fitting algorithm {\sc the payne} \citep{tin18c}. As shown in \citet{mar14} (e.g., their Figure 5), such disk stars are likely to be born cold, at least for $\tau \lesssim 7\,$Gyr \citep[but see also][for a different view]{bro04,bir13}. By restricting ourselves to these stars, we deem a scenario sensible, in which they were born ``cold'' (with low $J_z$) and were subsequently heated. We also restrict ourselves to stars with a Galactocentric height of $|z_{\rm GC}| < 0.5\,$kpc, as APOGEE is a low-latitude survey. As we will see, this makes the evaluation of $J_z$-selection function more straightforward. 

Applying all these criteria leaves a total of $20$,$764$ red clump stars. The full catalog used (after all these criteria are imposed) in this study, including their phase space actions, and stellar age properties is electronically available as Table~\ref{table1}, and a portion of which is shown in Table~\ref{table1} below.
 
As radial migration in the Galactic disk seems to be moderately strong \citep[see e.g.,][and references therein]{fra18}, the present-day Galactocentric radius, $R_{\rm GC}$ is unlikely to be the stars' birth radius, $R_{\rm birth}$ nor necessarily the mean Galactocentric distance at which they have experienced most of their heating; the latter is the quantity pertinent in building a global disk heating model. But global chemical evolution models including radial migration \citep{fra18} allow us to estimate $R_{\rm birth}$ from a star's age $\tau$ and metallicity [Fe/H]; this is effectively ``chemical-tagging'' in a weak sense \citep[e.g.,][]{fre02,tin15a}.

\begin{table*}
\begin{center}
\caption{APOGEE red clump value-added catalog. \label{table1}}
\begin{tabular}{lcccccccccccccccc}
\tableline \tableline
\\[-0.2cm]
APOGEE ID & [Fe/H]$_{\rm Payne}$ & [Mg/Fe]$_{\rm Payne}$ & Gaia Source ID & Gaia RA [$^\circ$] & Gaia Dec [$^\circ$] & Age $\tau$ [Gyr] & $d_{\rm phot}$ [kpc] \\[0.1cm]
\tableline
\\[-0.2cm]
2M00000446+5854329 & -0.115 & 0.044 & 422775384964691328 & 0.01860 & 58.90915 & 3.66 & 3.63 \\
2M00001071+6258172 & -0.447 & 0.131 & 430057759718424064 & 0.04468 & 62.97150 & 1.09 & 4.88 \\
2M00001199+6114138 &  0.007 & 0.148 & 429484398762416384 & 0.04996 & 61.23720 & 5.40 & 1.95 \\  
2M00001242+5524391 & -0.019 & 0.139 & 420487782302882560 & 0.05175 & 55.41084 & 4.59 & 3.87 \\ 
$\cdots$ & $\cdots$ & $\cdots$ & $\cdots$ & $\cdots$ & $\cdots$ & $\cdots$ & $\cdots$ \\
\tableline\\[0.2cm]

\tableline \tableline
\\[-0.2cm]
Apogee ID & $R_{\rm GC}$ [kpc] & $z_{\rm GC}$ [kpc] & $R_{\rm birth}$ [kpc] & $\overline{R}_{\rm GC}$ [kpc] & $J_R$ [kpc$\,$km/s] & $L_z$ [kpc$\,$km/s] & $J_z$ [kpc$\,$km/s] \\[0.1cm]
\tableline
\\[-0.2cm]
2M00000446+5854329 & 10.33 & -0.179 &   6.65 &   8.49 &   1.121 & 2462.3 & 0.957 \\
2M00001071+6258172 & 11.30 &  0.089 & 13.01 & 12.15 & 26.346 & 2347.2 & 0.472 \\
2M00001199+6114138 &   9.24 & -0.007 &   3.47 &   6.35 & 22.459 & 1936.2 & 6.615 \\
2M00001242+5524391 & 10.46 & -0.424 &   4.11 &   7.28 & 71.747 & 2086.9 & 4.361 \\
$\cdots$ & $\cdots$ & $\cdots$ & $\cdots$ & $\cdots$ & $\cdots$ & $\cdots$ & $\cdots$ \\
\tableline\\[0.2cm]
\end{tabular}
\\
{The original APOGEE red clump sample was derived in \citet{tin18a}, and here we provide a value-added version of that catalog, restricting to $\tau < 8\,$Gyr, $|z| < 0.5\,$kpc and the low-$\alpha$ stars. Column (1) shows the APOGEE ID; columns (2)-(3) show the APOGEE abundances measured using {\sc the payne} \citep{tin18c}; columns (4)-(6) indicate the Gaia source id, RA and Dec; column (7) shows the stellar ages inferred from the spectra via an asteroseismic training set; column (8) indicates the inferred photometric red clump distances from the Sun; columns (9)-(12) show the Galactocentric observed radii, Galactic heights, inferred birth radii and the mean radii, respectively; the mean radii are defined to be the average of the observed radii and the birth radii; columns (13)-(15) indicate the phase space actions (radial actions, angular momenta and vertical actions) of the stars.}
\end{center}
\end{table*}

\citet{fra18} has directly constrained such a metallicity--age--birth radius relation, [Fe/H]($R_{\rm birth},\tau)$, by simultaneously taking into account and marginalizing over the star formation history of the Milky Way and the radial migration of stars. We adopt their best-fitted [Fe/H]($R_{\rm birth},\tau)$ relation to infer $R_{\rm birth}({\rm [Fe/H]}, \tau)$, for our individual red clump stars. We then adopt for simplicity a presumed average radius (over the star's lifetime), $\overline{R}_{\rm GC}$ to be the average of the birth radius $R_{\rm birth}$ and the current radius $R_{\rm GC}$. For the model below we further assume that the star was effectively heated by the rate specified at $\overline{R}_{\rm GC}$. 

While this can be a crude estimate of the ``effective'' radius, to calculate the exact heating rate requires a complete knowledge of the exact trajectories of the stars. This is not possible due to the stochastic nature of orbits, either due to scattering and radial migration. Nonetheless, the typical migration scale length is $\sim$1.5 kpc \citep{fra18}, as we are probing a broad range of radii (3-14 kpc), a small change in the exact definition of "average" $\overline{R}_{\rm GC}$ would not change the result in this study qualitatively. We note that  the results presented in this study assume the empirical radial migration model derived in \citet{fra18}. We will defer the simultaneous fitting of both the radial migration and vertical heating to future studies.

Taking into account the uncertainties from the metallicity-age-radius relation, we assume the fractional uncertainty of $\overline{R}_{\rm GC}$ to be 10\% throughout this study (instead of $7\%$ uncertainty of $R_{\rm GC}$). We checked that assuming a larger or smaller uncertainty does not qualitatively change the result in this study.

In preparation of the modeling, we still have to calculate the vertical action $J_z$ for all stars, which we do with the {\sc galpy} package and its adopted Milky Way potential \citep{bov15b}. We assume $R_\odot = 8.2\,$kpc, $v_{\circ}(R_\odot) = 240\,$km/s \citep[e.g.,][]{sch12,bla16} and the solar motion with respect to the Local Standard of Rest to be $U=11.1\,$km/s, $V=12.24\,$km/s and $W=7.25\,$km/s \citep{sch12}. Throughout this study, we adopt the radial velocities from APOGEE DR14 and photometric distances derived above as they are more precise than Gaia's; Gaia DR2 only contributes to the proper motions of stars. But we note that a proper statistical inference which we will describe in the next section requires more than just the estimates of $J_z$, it also requires an estimate for the uncertainty of $J_z$, which is what we will derive next.

In the simple harmonic and epicycle approximations \citep[e.g.,][]{sol12}, one finds $J_z = E_z/\nu = (\frac{1}{2} v_z^2 + \frac{1}{2} \nu^2 z^2)/\nu = {v_{z,{\rm max}}}^2/2\nu$, where $E_z$ the vertical energy, $v_z$ is the vertical velocity and $\nu$ the vertical frequency. This implies that the uncertainty of $J_z$ can be approximated to be $\sigma_{\log Jz} = 2 \sigma_{\log vz}$. By definition, we have $v_z \sim \mu_z \cdot d$, where $\mu$ is the proper motion, it follows that $\sigma_{\log vz} = \sqrt{\sigma_{\log \mu z}^2 + \sigma_{\log d}^2} \simeq \sigma_{\log d}$. For photometric distances, we have $\sigma_{\log d}$ to be a constant, in our case, $7\%$. As a result, the vertical action uncertainty is only $\sim 15\%$ using red clump stars. We also tested that assuming a 10\% or 20\% error for $J_z$ does not change the results qualitatively. On the other hand, the uncertainty for $J_z$ can be significantly larger if one were to use the Gaia parallax distances. As $d \equiv 1/\varpi$, we have $\sigma_{\log d} = \sigma_{\log \varpi}$, But for Gaia, even at the bright end, only $\sigma_{\varpi}$ is constant, instead of $\sigma_{\log \varpi}$. It follows that $\sigma_{\log d} \sim d$, $\sigma_d \sim d^2$ and hence $\sigma_{\log Jz} \sim d \sim \sqrt{J_z}$ instead of a constant of $\sim 15\%$. As a consequence, the statistical inference would be significantly impeded by the large $J_z$ uncertainties for stars with $d \gtrsim 1\,$kpc \citep{cor18}. Therefore, with Gaia DR2 data alone, i.e., without the complementary spectroscopic data from APOGEE which allows us to obtain precise distances through the red clump sample, it would not be possible to constrain the vertical heating across the Milky Way.

Having described how we obtain the observable $\{ J_{z}, \overline{R}_{{\rm GC}}, \tau \}_i$, as well as their uncertainties, in the next section, we can now turn to describing the model which predicts the full distribution function of the vertical action $p(J_z|\overline{R}_{\rm GC},\tau)$. And subsequently, we will compare the data to the distribution through Bayesian inferences to obtain constraints on the model and to inform how the Milky Way disk was vertically heated to best describe the data.

%
%
%
%
%
%

\section{Vertical heating model and statistical inference}
\label{sec:model}

We now lay out simple parameterized, physically motivated models for the ``history'' of vertical stellar motions in the Galactic disk, and constrain it by these data. Such models should specify, as a function of Galactocentric radius, with what vertical action stars were born, and how this vertical action evolved subsequently. It is apparent from the outset that the present-day data, even if they were exhaustive and perfect, cannot fully constrain the past history of vertical motions, without further model restrictions or assumptions: after all, the solution that no vertical heating in the Galaxy ever happened and every star was born with its present day $J_z$ is mathematically viable and provides a perfect fit to the data. We therefore need to devise some physically plausible, though mathematically highly restricted model families as the context for any inference. We do this in three steps: first, we discuss how the assumption that the distribution of $J_z$ is quasi-isothermal simplifies the specification of a model. Second, we specify and formally introduce a family of models that assumes that stars were born with a certain (radius-dependent $J_z$) which then grew as some power of time; this models assumes implicitly that the observed $J_z-\tau$ relation reflects an evolutionary path, as $\Delta\tau$ and $-\Delta t$ are used near-interchangeably, where $\tau$ is the observed stellar age and $t$ is the time after the star was born. We note that at present time $t = \tau$, but at earlier time, these are two different quantities. For example, at a look-back time of $t_{\rm look-back} = 2\,$Gyr, a star that has a stellar age $\tau = 5\,$Gyr, would have $t=\tau-t_{\rm look-back} = 3\,$Gyr. Therefore, one should not confuse the measured $J_z - \tau$ relation from all stars with the ``evolutionary track'' $J_z -t$ relation for individual stellar populations. In fact, finally, in Section~\ref{sec:discussion}, we will presume that the heating mechanism is orbit scattering (with a $J_z\propto t^{0.5}$ time dependence for individual populations), and ask whether a density of scatterers that scaled with the presumed density of cold gas could explain the measured $J_z-\tau$ relation.

\begin{figure*}
\centering
\includegraphics[width=1.0\textwidth]{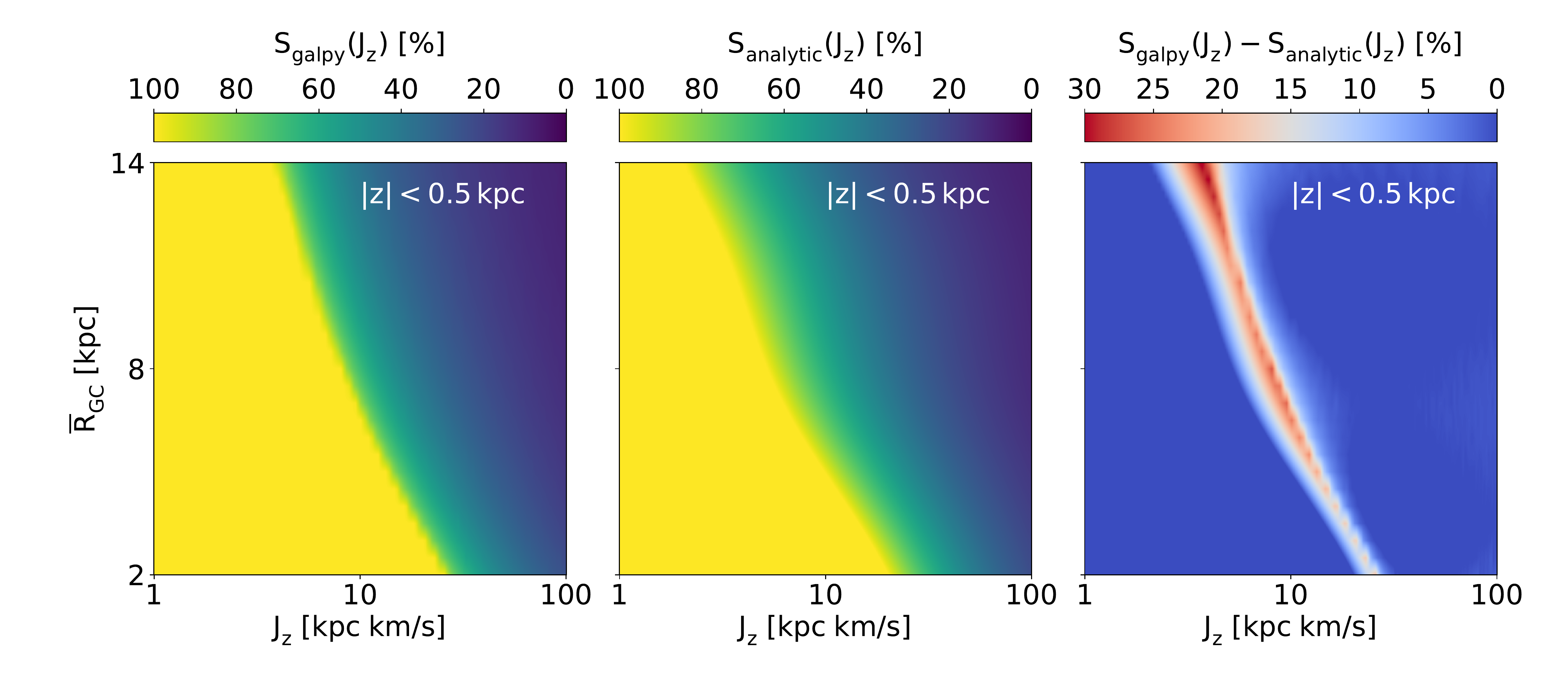}
\caption{The selection function S$(J_z|\overline{R}_{\rm GC})$ at different Galactocentric radii $\overline{R}_{\rm GC}$. The left panel shows the fraction in time, predicted by {\sc galpy}, for a star with vertical motion $J_z$ and radius $\overline{R}_{\rm GC}$ to spend in the region of $|z_{\rm GC}| < 0.5\,$kpc, i.e., how many stars are we statistically missing. At the inner disk with small $\overline{R}_{\rm GC}$, stars are spending more time at $|z_{\rm GC}| < 0.5\,$kpc, and the sample is more complete, because the inner disk has stronger restoring vertical force. At the outer disk, due to the weaker restoring force, the sample is severely biased toward stars with smaller vertical motions. Ignoring this bias could skew the inference on the Milky Way model. However, calculating the renormalization of the likelihood using a numerically evaluated selection function is computationally prohibitive. The middle panel shows an analytic approximation of the selection function adopted in this study (see Appendix for details). And the right panel shows the difference between the numerical selection function and the analytic approximation. At a given $\overline{R}_{\rm GC}$, the analytic approximation only does poorly at the transition point beyond which stars are not spending their entire time at $|z_{\rm GC}| < 0.5\,$kpc. We checked that such approximation errors which only affect a small portion of the stars do not adversely impact the conclusion of this study.}
\label{fig3}
\end{figure*}

%
%
%
%
%
%

\subsection{Isothermal distributions for $J_z$}

A star's current or past vertical action is clearly not a deterministic function of its birth and current radius and its age. But we can expect stars of such properties to have a probabilistic distribution of vertical actions that can be described by some characteristic quantities. An intuitive, and well-established model for  $p(J_z|\overline{R}_{\rm GC},\tau)$ would be an isothermal distribution, both at birth and subsequently (with a different temperature). For simplicity, we assume that we are in the harmonic limit, where the distribution can be written as:
\begin{equation}
p(J_z) \sim \exp(-\nu J_z/\sigma_z^2),
\end{equation}

\noindent
and the mean of the vertical action is $\widehat{J_z} = \sigma_z^2/\nu$. It follows that a normalized distribution of $J_z$ can be simply written as 
\begin{equation}
p(J_z|\overline{R}_{\rm GC},\tau) = \frac{1}{\widehat{J_z}(\overline{R}_{\rm GC},\tau)} \exp \left( -\frac{ J_z}{ \widehat{J_z}(\overline{R}_{\rm GC},\tau)} \right).
\end{equation}

\noindent
This {\it Ansatz} was proposed in \citet{bin10} and \citet{bin11}, and subsequently adopted in a number of studies such as \citet{tin13} and \citet{tri16}, as it seems to describe the observed action distribution of stellar disk sub-populations well. This {\it Ansatz} has the elegant property that the distribution is fully specified by the expectation value for the vertical action $\widehat{J_z}(\overline{R}_{\rm GC},\tau)$. As we will show, such an isothermal description through $\widehat{J_z}(\overline{R}_{\rm GC},\tau)$, describes the data well across all $\overline{R}_{\rm GC}$ and $\tau$. Specifying a model for the observed $\{J_z\}_i$ then reduces to specifying the set of possible $\widehat{J_z}(\overline{R}_{\rm GC},\tau)$.

%
%
%
%
%
%

\subsection{A parameterized model for the vertical action distribution as a function of $\tau$ and $\overline{R}_{\rm GC}$}

Following traditional approaches, we do not try to model the temporal evolution $\widehat{J_z}-t$ relation for individual populations, but model the global observed $J_z - \tau$ relation, i.e. $\widehat{J_z}(\overline{R}_{\rm GC},\tau)$, including here a $\overline{R}_{\rm GC}$-dependence, as a function of age $\tau$. Specifically, we adopt:
\begin{equation}
\label{eq-model-start}
\widehat{J_z}(\overline{R}_{\rm GC},\tau) \equiv \widehat{J_{z,0}} (\overline{R}_{\rm GC})\  +\  \DJz(\overline{R}_{\rm GC})\cdot \ageGyr^{\gammRC}.
\end{equation}

\noindent
Here, $\widehat{J_{z,0}} (\overline{R}_{\rm GC})$ is the stars' characteristic $J_z$ at birth, \DJz~ is typical increase in $J_z$ in the last Gyr, and $\gammRC$ specifies how the heating scales with age. This model makes two rather far-reaching assumptions: first, that $\widehat{J_{z,0}}$ is a distinct function of \Rbar , but has been constant with epoch or age $\tau$, at least for the last 8~Gyrs considered here. Second, that the time dependence of the vertical action (after birth) can be described as a power-law of exponent \gammRC , which changes with radius. To explore the $\overline{R}_{\rm GC}$-dependence of these terms we consider both a non-parametric approach (radial bins) and a polynomial in $\overline{R}_{\rm GC}$. These choices in model restrictions were guided by Occam's razor, i.e., the search for the simplest (or most rigid) model consistent with the data.

When considering individual radial bins with a spacing of $1\,$kpc, we constrain three parameters (as constant) $\widehat{J_{z,0}}$,\ \DJz~ and $\gamma$ in each. When constructing a global model for the radial dependence, we adopt a linear function for \DJz\ and \gammRC , and a cubic polynomial function for $\widehat{J_{z,0}}(\overline{R}_{\rm GC})$, a choice inspired by the radial dependences seen when fitting individual bins; to reduce covariances among the polynomial coefficients we take 8$\,$kpc as the pivot point of these functions, defining $\Delta R_{\rm GC}\equiv \overline{R}_{\rm GC} - 8\,{\rm kpc}$.

Taken together, the global model (specified by the model parameter vectors, ${\bf a},{\bf b},{\bf c}$) is then:
\begin{eqnarray}
\gammRC &\equiv& a_0 + a_1 \Delta R_{\rm GC},\\
\DJz (\overline{R}_{\rm GC}) &\equiv& b_0 + b_1 \Delta R_{\rm GC},\\
 \widehat{J_{,z,0}} (\overline{R}_{\rm GC}) &\equiv& 
 \sum_{j=0}^3 c_j {\Delta R_{\rm GC}}^j, \label{eq-model-end}.
\end{eqnarray}

%
%
%
%
%
%

\subsection{Selection function}

We aim to compare our model to stars from APOGEE \citep{maj17}, which is a sparse, low-latitude spectroscopic survey; at any Galactocentric radius $\overline{R}_{\mathrm GC}$, the probability of a star entering this sample depends on its height $z$ above the midplane, and therefore on its $J_z$, as stars with large $J_z$ spend a larger fraction of their orbit well away from the mid-plane. For a data-model comparison, we need to construct an approximate selection function $S(J_z|\overline{R}_{\rm GC},\tau)$. Constructing a selection function would be significantly more involved were we to aim for quantifying the full $p(J_{z},\overline{R}_{\rm GC},\tau)$: the radial selection (even more than the $z$-selection) function arises from a combination of the APOGEE targeting strategy and dust extinction; and red clump stars are strongly biased against old stars and towards $\sim 2$~Gyr old stars. Here we circumvent this problem by sampling our model posterior from $p(J_{z}|\overline{R}_{\rm GC},\tau)$, focusing on the evolution of $J_z$ {\em at any given radius and time}. 

Fortunately, as shown in Fig.~\ref{fig2}, an approximate selection function $S(J_z|\overline{R}_{\rm GC})$ can be quite straightforwardly calculated from simple orbit integration over a presumed Milky Way potential; it has approximation that enables an analytically integrable normalization of the likelihood function, as we describe in the Appendix (also see Fig.~\ref{fig3}).

%
%
%
%
%
%

\subsection{Data likelihood and model posteriors}

Finally, when the selection function is included, instead of directly comparing the data to $p(J_z|\overline{R}_{\rm GC},\tau,{\bf p})$ predicted by the isothermal distribution function, the model prediction should be revised to be
\begin{equation}
\widetilde{p}(J_z|\overline{R}_{\rm GC},\tau,{\bf p}) = \frac{1}{C(\overline{R}_{\rm GC},\tau,{\bf{p}})} p(J_z|\overline{R}_{\rm GC},\tau,{\bf{p}}) \cdot S(J_z | \overline{R}_{\rm GC})
\end{equation}

\noindent
where $C(\overline{R}_{\rm GC},\tau,{\bf{p}})$ is a normalization (see Appendix).

With this in mind, given the data $\{J_z, \overline{R}_{\rm GC}, \tau\}_i$ of individual stars in the red clump sample, the global model parameters ${\bf p}\equiv (a_0, a_1, b_0,b_1, c_0, c_1, c_2, c_3)$ can then be inferred through Bayesian inference;  in the case of fitting individual radial bins, an analogous procedure applies to $\widehat{J_{z,0}}$,\ \DJz~ and $\gamma$. 

More precisely, assuming an uniform prior, we have the posterior of {\bf p} to be
\begin{eqnarray}
&&p({\bf p}| \{ J_{z}, \overline{R}_{\rm GC}, \tau\}) = \prod_{i=1} \widetilde{p}(J_{z,i}| \overline{R}_{{\rm GC},i}, \tau_{i}, {\bf p}) \nonumber \\
&=& \prod_{i=1}  \int {\rm d} J_{{\rm true},i} \int {\rm d} \overline{R}_{\rm true,i} \int {\rm d}\tau_{\rm true,i} \;  p(J_{z,i}| J_{z,i,{\rm true}}) \nonumber \\
&&\;\;  \widetilde{p}(J_{z,{\rm true}}| \overline{R}_{\rm true,i}, \tau_{\rm true,i} , {\bf p}) \cdot p(\overline{R}_{\rm true,i}|\overline{R}_{{\rm GC,}i}) \cdot p(\tau_{\rm true,i}|\tau_i)
\end{eqnarray}

\noindent
where $J_{z,i,{\rm true}}$, $\overline{R}_{\rm true}$ and $\tau_{\rm true}$ reflect the ``noiseless'' observable, as opposed to the observed $J_{z,i}$, $\overline{R}_{{\rm GC},i}$, $\tau_i$ which are subjected to measurement uncertainties. In practice, the integrals were calculated via Monte Carlo sampling 10 independent realizations of the true observable, assuming $\sigma_{Jz}/J_z = 15\%$, $\sigma_{\overline{R}_{\rm GC}}/\overline{R}_{\rm GC} = 10\%,$ and $\sigma_{\tau}/\tau = 25\%$, essentially Monte Carlo integrating over the measurement uncertainties of the observable.

\begin{figure*}
\centering
\includegraphics[width=1.0\textwidth]{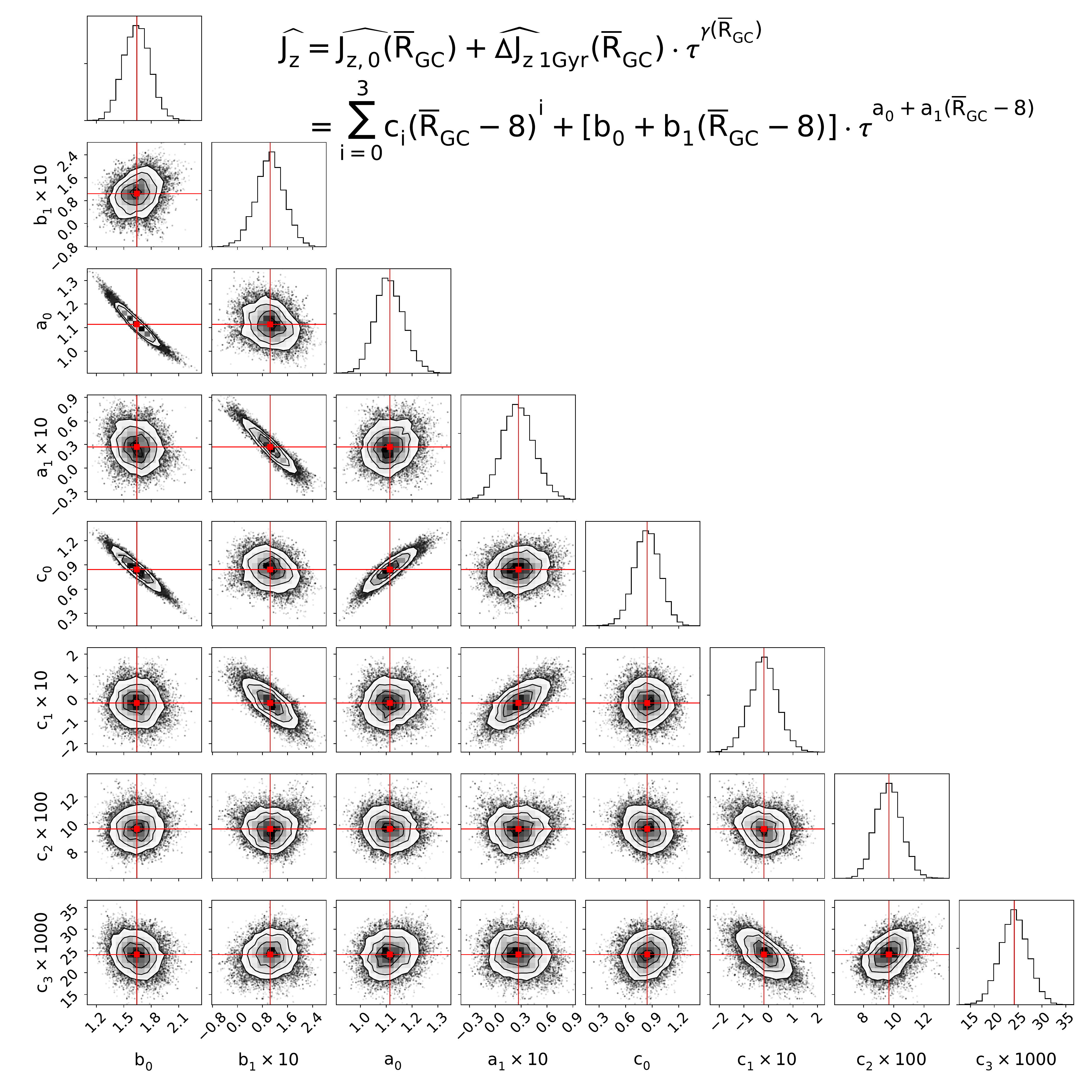}
\caption{The posterior of model parameters depicting the vertical heating and the birth temperature of the Milky Way thin disk. We assume the stellar vertical action distribution at any given Galactocentric radius $\overline{R}_{\rm GC}$ and stellar age $\tau$ can be approximated by an isothermal distribution $p(J_z|\overline{R}_{\rm GC},\tau) = \exp(-J_z/\widehat{J_z}(\overline{R}_{\rm GC},\tau))$, where $\widehat{J_z} = \widehat{J_{z,0}}(\overline{R}_{\rm GC}) + \DJz (\overline{R}_{\rm GC}) \cdot \tau^{\gamma(\overline{R}_{\rm GC})}$. The first two parameters $b_0$ and $b_1$ constrain the intercept and slope of $\DJz (\overline{R}_{\rm GC})$. The next two parameters $a_0$ and $a_1$ shows the intercept and slope of the power law index $\gamma (\overline{R}_{\rm GC})$, and the last four parameters fit for the cubic radial dependence of the birth temperature of stars, $\widehat{J_{z,0}}(\overline{R}_{\rm GC})$.}
\label{fig4}
\end{figure*}

\begin{figure*}
\centering
\includegraphics[width=1.0\textwidth]{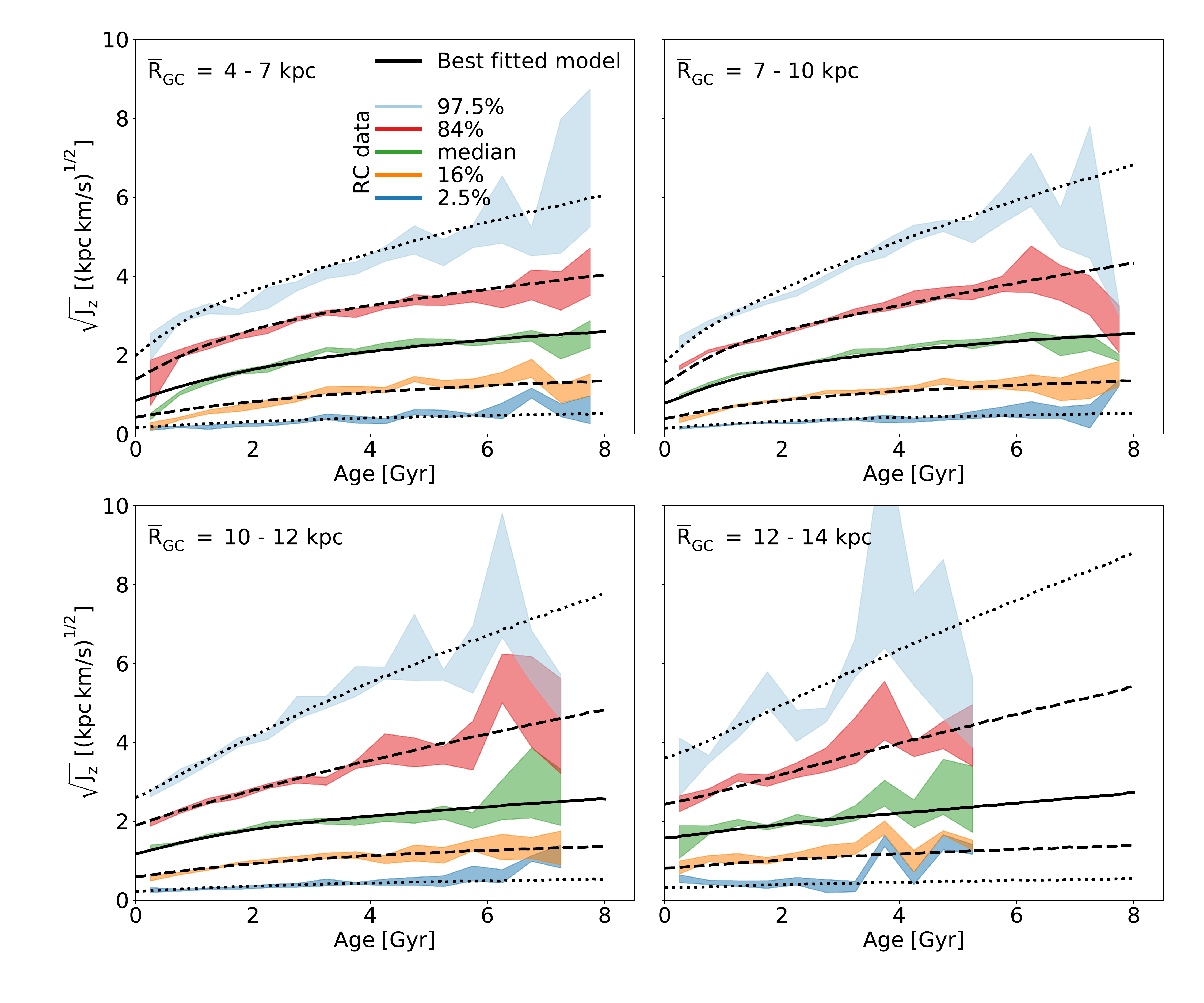}
\caption{Selection function corrected vertical action distribution $p(J_z|\overline{R}_{\rm GC},\tau)$ at different Galactocentric radii (different panels) as a function of stellar ages. In each panel, we plot the predictions from the best-fitted vertical heating model in black lines (2.5, 16, 50, 84, 97.5 percentiles, respectively), evaluated at the median $\overline{R}_{\rm GC}$ of the observed sample for that panel. When evaluating the predictions, we also take into account the selection function $S(J_z|\overline{R}_{\rm GC},\tau)$ and the measurement uncertainties of $J_z$ ($15\%$). The overplotted shaded regions in colors show the observations from our APOGEE red clump sample. We only show results for spatial-temporal bins with more than 5 stars. The uncertainties of the observations are estimated by bootstrapping the data 1000 times. The model shows an excellent agreement with the data. Remarkably, the ``natural width'' of $p(J_z|\overline{R}_{\rm GC},\tau)$ predicted by the isothermal distribution functions captures the moments of $p(J_z|\overline{R}_{\rm GC},\tau)$ very well without the need of additional scatter term.}
\label{fig5} 
\end{figure*}

\begin{figure*}
\centering
\includegraphics[width=1.0\textwidth]{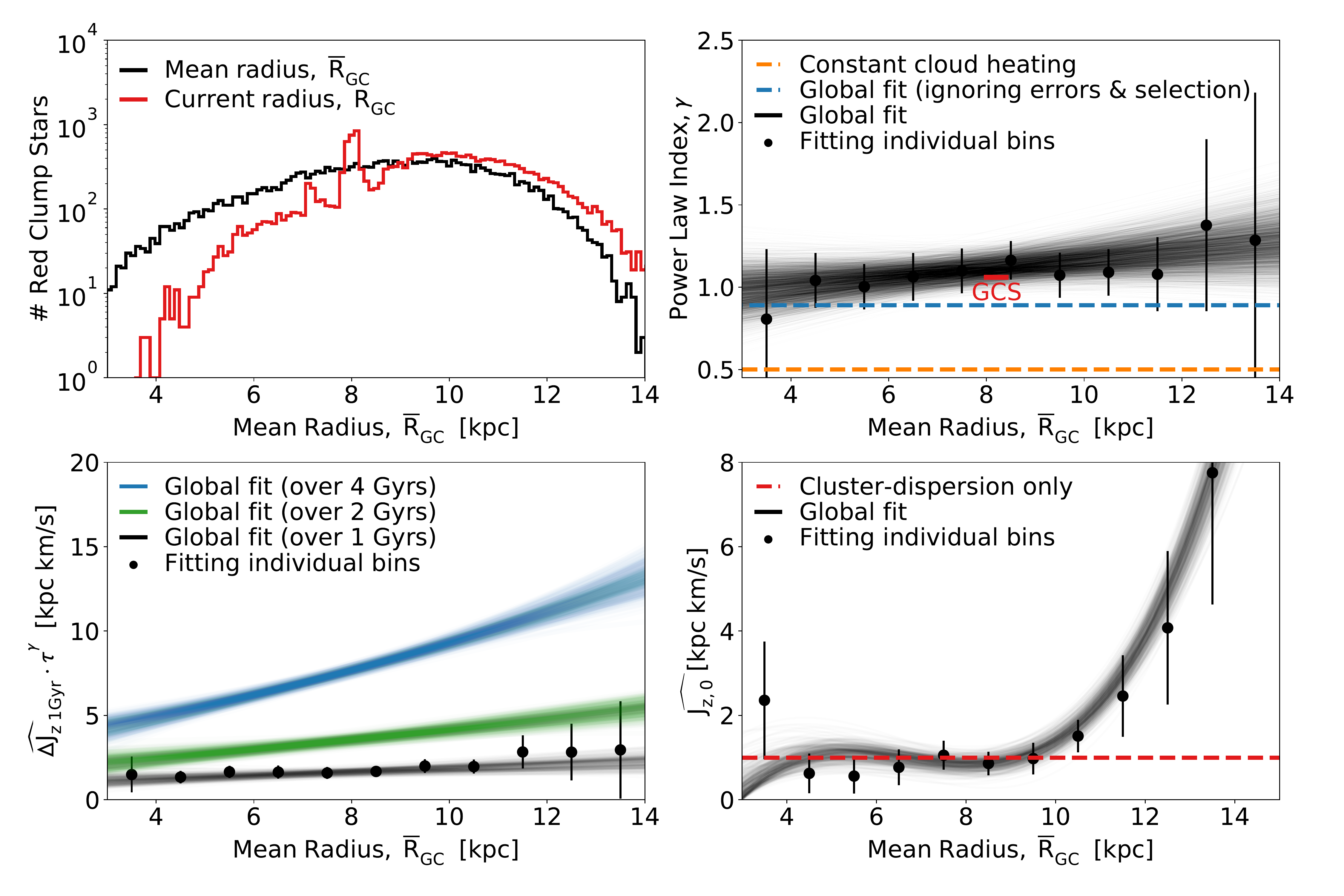}
\caption{Radial dependence of the vertical heating and birth temperature of stars. The top left panel shows the radial distribution of the APOGEE red clump sample adopted in this study. The red histogram illustrates the current Galactocentric radii and the black histogram the average of the current and estimated birth radii, $\overline{R}_{\rm GC}$. For the other panels, we show the MCMC posterior of our vertical heating model, taking into account the posterior covariances. The black points show the mean and $1\sigma$ of the ``non-parametric'' fit of individual bins without assuming any radial dependence, and the black lines the global fit. The top right panel shows the temporal power law index. The orange line shows the expected power law if the vertical heating of the young disk is purely due to GMC scattering {\em with a time-independent scattering amplitude}. The blue line shows our global fit if we do not take into account the measurement uncertainties and the selection function of $J_z$. When taking into account the selection function, our estimates are consistent with the GCS value near the solar neighborhood  (shown in red) and show a linear radial grow in the power law index. The radially linear growth effect is also manifested in the bottom left panel which shows the total vertical heating that is not accounted for by the birth temperature. The black, green and blue lines indicate the amount of vertical disk heating at 1, 2 and 4$\,$Gyrs respective. Finally, the bottom right panel demonstrates the vertical birth actions of stars. For stars in the disk with $\overline{R}_{\rm GC} < 10\,$kpc, the birth action is consistent with the expected value from a ISM gas dispersion velocity of $7\,$km/s, which is illustrated in the red dashed line. The best-fitted model show that stars beyond $10\,$kpc from the Galactic center were born significantly hotter, indicating a waning disk self-gravity.}
\label{fig6}
\end{figure*}

%
%
%
%
%
%

\section{Results: a global view of $\widehat{J_z}(\overline{R}_{\rm GC},\tau)$}
\label{sec:results}
 
Following the model and its corresponding likelihood as described in the previous section, we sample the posterior of our model parameters via MCMC using the {\sc emcee} package \citep{for13}. We first fit for individual radial bins where we bin our data into 1 kpc radial bins, spanning from $\overline{R}_{\rm GC} = 3\,$kpc to $14\,$kpc. We fit for three parameters $\widehat{J_{z,0}}$,~\DJz~ and $\gamma$ for each bin. Noting that a small fraction of our sample could be contamination from the thick disk or the halo stars, as well as stars older than $8\,$Gyr but were mistakenly inferred to be younger, after the first round of fitting, we evaluate the likelihood for individual objects given the best parameter and cull sample with log-likelihood $\ln \mathcal{L} < -6$. We tested that this choice gives the largest improvement to the comparison of the data to the model, But we note this criterion only discards $1.3 \%$ of the data, and without this cut, the model still gives a good prediction of the data. We subsequently refit the results using the ``clean'' sample which defines the final results for the individual bins. Finally, we combine all the sample from the individual bin fitting, and fit for the global model where we fit for the radial dependence of $\widehat{J_{z,0}}$,~\DJz~ and $\gamma$  simultaneously, assuming the functional form presented in Eq.~(\ref{eq-model-start})--(\ref{eq-model-end}).
\begin{table}
\begin{center}
\caption{Best-fitted parameters for the vertical heating model (Eq.~\ref{eq-model-start}-\ref{eq-model-end})\label{table2}}
\begin{tabular}{cc}
\tableline \tableline
\\[-0.2cm]
Parameter & Best-fitted value \\[0.1cm]
\tableline
\\[-0.2cm]
$a_0$ & 1.09 \\
$a_1$ & 0.06 \\
$b_0$ & 1.81  \\
$b_1$ & 0.05 \\
$c_0$ & 0.91 \\
$c_1$ & 0.18 \\
$c_2$ & 0.087 \\
$c_3$ & 0.014 \\[0.1cm]
\tableline\\[0.2cm]
\end{tabular}
\end{center}
\end{table}

Fig.~\ref{fig4} shows the sampling of the model parameter posterior. At each panel, the crosshair indicates the best-fitted model, defined through the mean of the marginalized distribution. Choosing a pivot point of $\overline{R}_{\rm GC} = 8\,{\rm kpc}$\footnote{The pivot point is chosen for numerical convenience to reduce the covariances of the posterior; it should not be confused with the Solar radius $R_\odot$, which is at 8.2 kpc.} as explained in Section~\ref{sec:model} reduces a large part of the posterior covariances. But some covariances persist because stars can either be born at a slightly higher birth temperature or could be slightly more heated later on. But given a broad range of stellar ages probed in this study, especially the youngest stars set the possible acceptable range for the birth temperature, the model parameters are not entirely degenerate, and the MCMC chain does converge to the best-fitted model as illustrated in Fig.~\ref{fig4}, and we find the best-fitted isothermal model with $p(J_z) \sim \exp(-J_z/\widehat{J_z})$ to be (the best-fitted parameters are also listed in Table~\ref{table2})
\begin{eqnarray}
\widehat{J_z}(\overline{R}_{\rm GC},\tau) &=& \widehat{J_{z,0}} (\overline{R}_{\rm GC}) + \DJz (\overline{R}_{\rm GC}) \ageGyr^{\gammRC}  \nonumber \\
&=& \Big(0.91 + 0.18 \, \Delta R_{\rm GC} + 0.087 \, \Delta R_{\rm GC}^2 \nonumber \\
&&\quad+ 0.014 \, \Delta R_{\rm GC}^3 \Big) \nonumber \\
&&\; +  \Big(1.81 + 0.050 \, \Delta R_{\rm GC}\Big) \cdot \tau^{1.09 + 0.060\, \Delta R_{\rm GC}}.
\label{eq:best-fitted-model}
\end{eqnarray}

To illustrate that the model from Eq.~\ref{eq-model-start} is indeed an adequate representation of the data, in Fig.~\ref{fig5} we compare the predictions of $J_z$ at different radii and stellar ages from the best-fitted model to the data. When projecting the predictions, we also take into account the selection function and the uncertainty in $J_z$. In particular, we sample from the best-fitted model, but down weight the mock sample by the selection function $S(J_z|\overline{R}_{\rm GC})$ evaluated with the mock sample values in $J_z$ and $\overline{R}_{\rm GC}$. We also perturb the $J_z$ assuming the same $J_z$ uncertainties of the data. The black lines demonstrate predictions from the model, and the corresponding shaded regions in colors, where the black lines center, demonstrate the data. For the data, the uncertainties as shown in the shaded regions are evaluated by 1000-fold bootstrapping. As for the model predictions, the solid line illustrates the median of the prediction and the dashed and dotted lines show the $\pm1\sigma$ and $\pm 2\sigma$ ranges. Different panels illustrate the evolution of the vertical action at different Galactocentric radii. Note that, we plot the $y$-axis in $\sqrt{J_z}$ instead of $J_z$ to show the full dynamical range of $J_z$. 

It is clear from the bottom panels of Fig.~\ref{fig5} that stars with $\overline{R}_{\rm GC} > 10\,$kpc have higher birth actions, even though the selection function should bias against hot stars. The vertical heating rate also increases visibly at large Galactocentric radii, and we will quantify that in more details below. Importantly, the model agrees very well with the data in the sense that the model predicts not only the expectation of $J_z$ but also the full distribution. This is remarkable, as for an isothermal $J_z$-distribution the width or ``scatter'' is uniquely given by the expectation value $\widehat{J_z}$ without any other parameters. Finally, we caution that the median values here should not be directly compared to  Eq~(\ref{eq:best-fitted-model}) (the mean of the distribution) because (a) the median of an exponential profile is $\ln 2$ times of the mean, and (b) the model predictions here are weighted by the selection function.

Fig.~\ref{fig6} further shows the resulting posteriors for the various components of the vertical heating model as a function of $\overline{R}_{\rm GC}$. When evaluating the model predictions, we draw directly from the posterior chain, in other words, the model predictions take into account the covariances of the posterior. The top panel shows the radial distribution of our data. The red histogram illustrates the current radii of the stars, and the black histogram shows the average of the estimated birth radii and the current radii of the stars. As shown, our APOGEE red clump sample covers a considerable fraction of the Milky Way disk, from $\overline{R}_{\rm GC} = 3\,$kpc to $14\,$kpc. Although there are a few stars that are located beyond this range, the number of stars per radial bin is too small ($<10$) for any reliable statistical inference beyond this range. Hence we only consider stars that are $3\, \leq \overline{R}_{\rm GC} \leq 14\,$kpc in this study. The panel also shows that the average radii are generally smaller than the current radii, demonstrating that there is an overall outward radial migration of stars. 

For the other panels, the black solid points show the mean and the $1\sigma$ standard deviation of the posterior when we fit a ``non-parametric'' model, i.e., by fitting data of different individual radial bins separately. We first fit data of individual radial bins to make sure that when we fit for the global model, the global model does not smooth out interesting features or create spurious features due to overfitting. The black lines in these panels show the global fit, where we assume a cubic radial dependence for the birth action $\widehat{J_{z,0}}(\overline{R}_{\rm GC})$, and a linear model for the heating rate $\DJz (\overline{R}_{\rm GC})$ and the temporal power law index $\gamma(\overline{R}_{\rm GC})$. As shown in Fig.~\ref{fig6}, the radial trend probed by the global fits agrees with the non-parametric version. 

In the following, we will only focus on the results from the global fit and will discuss how these parameters vary as a function of $\overline{R}_{\rm GC}$, as this analysis presents the first comprehensive view of the age-dependent vertical actions of Galactic disk stars beyond the Solar neighborhood. We iterate that the results in this study assume the radial migration models from \citet{fra18}, a more robust approach for future studies would be fitting both the radial migration and vertical heating simultaneously.

First, we focus on the inferred birth action of the stars as a function of $\overline{R}_{\rm GC}$. We plot the posterior of $\widehat{J_{z,0}} (\overline{R}_{\rm GC})$ in the bottom right panel of Fig.~\ref{fig6}. This birth temperature information primarily comes from the youngest population of stars in our sample; it is a well-constrained quantity, albeit it is slightly correlated with the vertical heating of the disk (see Fig.~\ref{fig4}). The figure shows that even though we assume that stars were born cold, at all radii, the birth action is non-zero. This is not surprising because the ISM in the midplane has a finite velocity dispersion. Furthermore, it is also known that star clusters disperse with a non-zero finite velocity. Therefore, even if stars were born cold, they would have a finite ``zero-point'' $\widehat{J_{z,0}}$ which expectation we will estimate next. 

We can compare $\widehat{J_{z,0}}$ to known cold ISM properties, the material from which these stars were just born, using the ISM velocity dispersion and the scale-height. The vertical action is approximately $\widehat{J_z} = E_z/\nu \simeq \sigma_z^2/\nu$, and in the isothermal limit we have $\sigma_z = \sqrt{2 \pi G \rho} h_z$. Note that, $\nu$ could be approximated by assuming that stars are oscillating in a uniform density medium. By solving the Poisson equation, we arrive at $\ddot{z} = -4 \pi G \rho z$ which is a simple harmonic equation with the vertical oscillation frequency $\nu = \sqrt{4 \pi G \rho}$. Substituting the formula for $\sigma_z$, it follows that $\nu =\sqrt{2} \sigma_z/h_z$, and finally substituting this to the $\widehat{J_z}$ formula, we find $\widehat{J_{z,0}} \simeq \sigma_z h_z / \sqrt{2}$ which we will adopt to estimate the zero-point vertical action. More quantitatively, if we assume the initial velocity dispersion of the ISM or the star clusters to be $7\,$km/s \citep[e.g.,][]{sta89,sta05,sta06,aum09,aum17}, and assuming the molecular gas scale height to be $0.2\,$kpc \citep[e.g.,][]{mara17}, we find $\widehat{J_{z,0}} = 1.0\;{\rm kpc}\;{\rm km/s}$, which is plotted as the red dashed line in the bottom right panel. Equivalently, the derived $J_{z,0}$ might imply that the average density in the vertical column where the stars oscillate is about half of the Solar neighborhood density $\rho_0$; if $\rho = \rho_0/2 \simeq 0.05 {\rm M}_\odot {\rm pc}^{-3}$, we have $\nu = \sqrt{4 \pi G \rho} = 50 \;{\rm Gyr}^{-1}$, and hence $J_{z,0} = \sigma_z^2/\nu = 1 \;{\rm kpc}\;{\rm km/s}$ as observed.

Fig.~\ref{fig6} demonstrates that for the disk stars born  at $\overline{R}_{\rm GC}\lesssim 10\;$kpc, the birth action is consistent with the velocity dispersion and scale height of the ISM and young clusters. Interestingly, the model fit shows that stars with $\overline{R}_{\rm GC} > 10\,$kpc were born with increasingly larger vertical motions: $\widehat{J_{z,0}} $ is reaching $ \simeq 8 \,{\rm kpc}\,{\rm km/s}$ at $\overline{R}_{\rm GC} = 14\,$kpc, almost an order of magnitude higher than in the inner disk. Such a strong radial dependence of birth temperature may reflect that waning disk self-gravity  facilitates the flaring and warping of the birth ISM \citep{nak03,amo17}. While the Galactic warp is not explicitly included in the model, its effect could manifest itself through the birth temperature of the stars -- stars born at a higher height would inherit a larger orbit oscillation and hence a larger initial vertical action. Flaring is another possibility. As $h_z \sim \sigma_z/\sqrt{\rho} \sim \sqrt{J_z/ \sqrt{\rho}}$, we have $J_z \sim h_z^2 \sqrt{\rho}$. At the other region, if $h_z$ increases due to flare, but at the same time, if $\rho$ only decreases gradually (e.g., from dark matter halo contribution),  a higher $J_{z,0}$ is expected as a result.
\begin{figure*}
\centering
\includegraphics[width=1.0\textwidth]{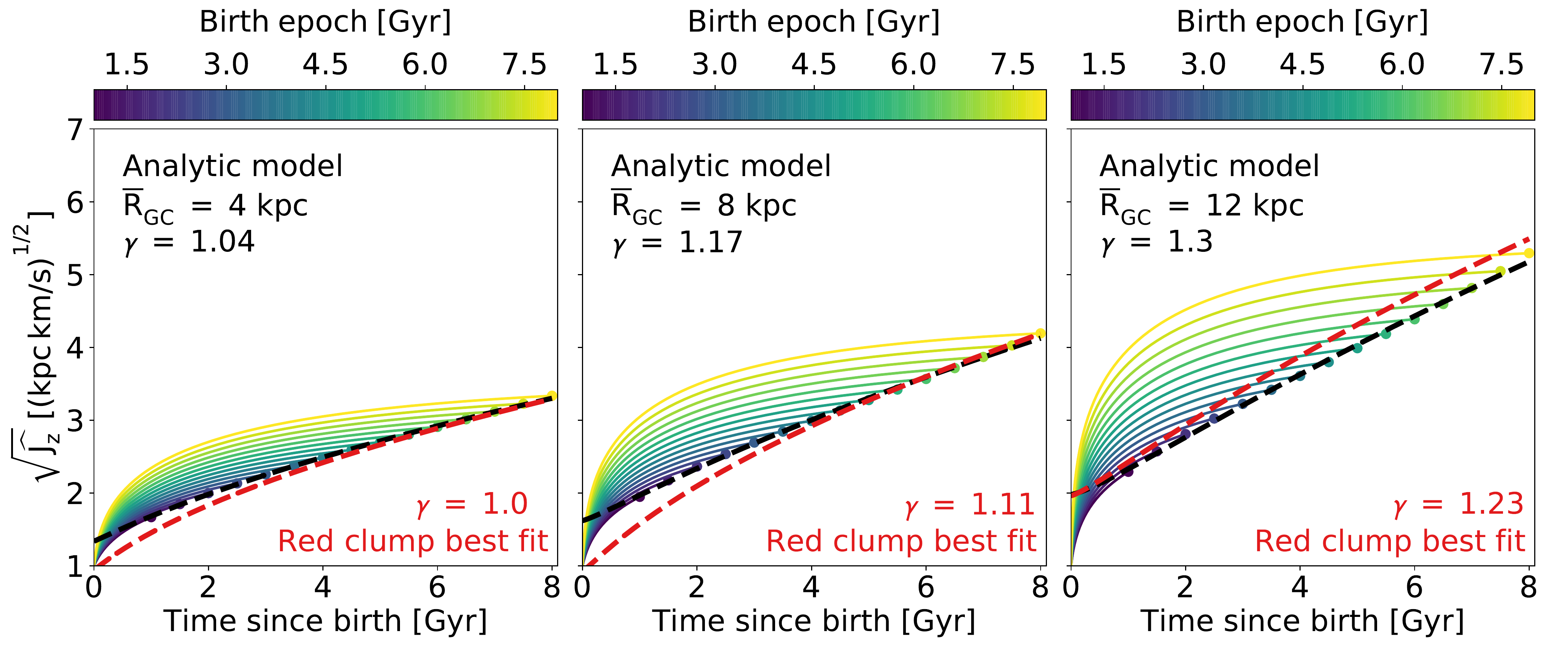}
\caption{An analytic model which explains the observed $\widehat{J_z} \sim \tau^{\sim 1}$ relation and radial growth in the $\gamma$ power law index. We consider an analytic model where individual stellar populations get heated only by GMC scattering. Individual lines in color show the evolution of various stellar populations formed at different times, each track ending at the present epoch. We assume that the scattering amplitude is proportional to $\Sigma_{\rm SFR}^{1/\alpha_{\rm KS}} (\overline{R}_{GC}, \tau)$ because a higher star-formation rate entails a higher GMC concentration, and the scattering amplitude is inversely proportional to the stellar density $\Sigma_*(\overline{R}_{GC}, \tau)$ due to a higher restoring force from the stellar disk. We also assume that the stellar disk has gone through an inside-out growth. We assume the initial condition $\widehat{J_z}(t=0) = 1\;$kpc$\,$km/s, consistent with the vertical action set by the ISM dispersion. The black dashed lines demonstrate the best-fitted power-law fit of the mock data created through this model. And the red lines indicate the best red clump global fit presented in this study. From left to right, we show the evolution at different radial bins. Despite its simplicity, the model reproduces the observed trend remarkably well, indicating that $\gamma$ and its radial growth, could be explained solely by the evolution of the Milky Way with the GMC scattering being the dominant source for vertical heating.}
\label{fig7}
\end{figure*}

But this model also describes the change of $\widehat{J_z}$ with age $\tau$. In this model, this is quantified by $\DJz (\overline{R}_{\rm GC}) \cdot \tau^{\gamma (\overline{R}_{\rm GC})}$. The black lines in the bottom left panel of Fig.~\ref{fig6} shows the posterior of $\DJz (\overline{R}_{\rm GC})$ which describes the apparent heating over the last Gyr. The posterior shows that the short-term heating rate is mild, $\DJz (\overline{R}_{\rm GC}) \approx 1.6 \, {\rm kpc\;km/s\;Gyr^{-1}}$ and with little radial dependence. How does this value compares to the GCS measurement? Assuming the average density to be $\rho_0 = 0.05 M_\odot {\rm pc}^{-3}$ as we have derived above via the birth temperature $J_{z,0}$, we have the local vertical frequency to be $\nu = 50 \; {\rm Gyr}^{-1}$. Recall that \citet{hol09} (fig 7) finds ${\rm d} \sigma_z^2/{\rm d} t \simeq 80 {\rm km/s}$ which implies $\DJz (\overline{R}_{\rm GC}) = \frac{1}{\nu} {\rm d} \sigma_z^2/{\rm d} t \approx 1.6 \, {\rm kpc\;km/s\;Gyr^{-1}}$, consistent with the value we observe.

The long-term change of the actions with age is described by the power law index $\gamma(\overline{R}_{\rm GC})$, plotted in the top right panel of Fig.~\ref{fig6}. The best-fitted model shows a linear growth with age, $\gamma \simeq 1$, which is also consistent with the Solar neighborhood measurement $\sigma_z \sim \tau^{0.53}$ \citep[][with $\widehat{J_z} \sim \sigma_z^2$ in the harmonic limit]{hol09}, which is overplotted as a red symbol in the top right panel. If we were to neglect the measurement uncertainties and selection function in our study, we would get a lower heating rate, illustrated by the blue dashed line. But as opposed to the local measurements, when both the selection function and the measurement uncertainties are included, the best fit shows $\gamma$ rising from about $1.0$ at $\overline{R}_{\rm GC} = 3\,{\rm kpc}$, to $\gamma = 1.3$ at $\overline{R}_{\rm GC} = 14\,{\rm kpc}$. The stronger long term heating for the outer disk is also demonstrated in the bottom left panel where the green and blue lines show the vertical heating of stars at different $\overline{R}_{\rm GC}$, over 2 Gyrs and 4 Gyrs, respectively.

The value of $\gamma$ and its radial dependence are noteworthy in two respects, First, our finding shows that the stellar age dependence of vertical motions differs markedly (with $\gamma \gtrsim 1$) throughout the disk from the expectation of the {\it temporal evolution} of constant gradual orbit heating by, say, scattering, which should lead to $\widehat{J_z} \sim \tau^{0.5}$ (orange line in the top right panel). Second, $\gamma$ raises outward, implying at face value a different heating history or mechanism.

This stark discrepancy of the observed age-scaling of $\widehat{J_z}(\overline{R}_{\rm GC},\tau)$, now seen at all radii, with the expectation of vertical orbit heating by a given population of scatters near the disk plane, $\widehat{J_z}\propto \tau^{0.5}$, leaves us with several alternative explanations: either there is a very different heating mechanism (satellites etc.), or the stars' birth temperature, $\widehat{J_{z,0}}$, evolved strongly with time, or orbit scattering is the heating mechanism, but its efficacy evolved strongly with time, as suggested e.g., by \citet{aum16}.

In the next Section we lay out a simple, physically motivated model for epoch-dependent orbit heating by scattering (e.g., off GMC's) that explains, at least qualitatively, both $\gamma\sim 1$ and its increase with  $\overline{R}_{\rm GC}$.

%
%
%
%
%
%

\section{Discussion: A simple orbit scattering for heating disk stars}
\label{sec:discussion}

It may be tempting to identify the age-dependent increase of vertical actions, $\widehat{J_z}(\tau)$, with an evolutionary heating path. But really, $\widehat{J_z}(\overline{R}_{\rm GC},\tau)$ simply reflects the combination of the birth action and the integrated subsequent heating that stars of current age $\tau$ at $\overline{R}_{\rm GC}$, i.e., the present-day end-points of possibly different, age-dependent evolutionary paths. When viewed just mathematically, success in matching these data with different combinations of birth actions and evolving scatterers would seem unsurprising, perhaps almost trivial. But here we show that an approximate match to the data can be achieved if we take one particular, astrophysically plausible evolution for the scattering strengths.

The Milky Way's SFR has been declining from high redshift to the present, and star-formation has proceeded inside out, building up the disk surface mass density at a radius-dependent rate. The rate of the resulting change in a star's vertical action will depend on both density of scatterers and the vertical orbit frequency, which in turn depends on the local disk mass density. These factors can hardly be constant with cosmic epoch.  ``Vertical heating tracks'' will then be different from the present-day $J_z - \tau$ relation. This is qualitatively illustrated in  Fig.~\ref{fig7} \citep[see also][]{aum16}, where individual lines in color indicate the evolution of individual populations born at different times, all following roughly $\widehat{J_z} \sim t^{0.5}$ but with different scattering amplitudes. Each line ends with the only observable prediction at the current epoch; the ensemble of evolutionary end points can be described as a power-law with stellar age $\tau$, which will have a different power-law index $\gamma$.

In the following we present a simple analytic model that quantifies these effects. In this model we denote the present-day age of the stars as $\tau$, the time since their birth as $t$, which implies a look-back time $t_{\rm look-back}\equiv \tau - t$. We take external pieces of information to estimate how the density of scatterers, denoted as $\Sigma_{\rm GMC} (\overline{R}_{\rm GC},\tau)$ and the vertical frequency, $\nu(\overline{R}_{\rm GC},\tau) $ (or restoring force) may have evolved. 

We start out with the basic equation describing vertical heating by orbit scattering, \citep[p.123][]{bin08} 
\begin{equation}
\label{eq:vertical-energy}
\frac{{\rm d}E_z}{{\rm d}t} \sim \frac{\Sigma_{\rm GMC}}{E_z^\delta},
\end{equation}
where $\delta\approx 0.5$ if the layer of scatterers is as thick as the vertical oscillations of the stars, and $\delta=1$, if the layer of scatterers is thin; numerical simulations suggest $\delta\approx 1$ to be the applicable regime. We want to cast this equation in terms of $J_z$ and allow for time-dependences. If we assume that we are in the limit of harmonic oscillations, and that the midplane density is dominated by the stellar mass and is uniform as a function of height, we have $J_z\equiv E_z/\nu$ and $\nu \propto \sqrt\Sigma_* (\overline{R}_{\rm GC},\tau)$. 

One then gets
\begin{equation}
\nu(\overline{R}_{\rm GC},\tau)\frac{{\rm d}\widehat{J_z}}{{\rm d}t} (\overline{R}_{\rm GC},\tau)
\sim \frac{\Sigma_{\rm GMC}(\overline{R}_{\rm GC},\tau)}{\widehat{J_z}(\overline{R}_{\rm GC},\tau)\nu(\overline{R}_{\rm GC},\tau)}, 
\end{equation}

\noindent
or with a scaling factor $C$ as free parameter,
\begin{equation}
\frac{{\rm d}\widehat{J_z}}{{\rm d}t} (\overline{R}_{\rm GC},\tau)
= C \times \frac{\Sigma_{\rm GMC}(\overline{R}_{\rm GC},\tau)}{\widehat{J_z}(\overline{R}_{\rm GC},\tau)\nu^2(\overline{R}_{\rm GC},\tau)}.
\end{equation}

\noindent
Since $\Sigma_* \sim \nu^2$,
\begin{equation}
\label{eq:heating_equation}
\frac{{\rm d}\widehat{J_z}}{{\rm d}t} (\overline{R}_{\rm GC},\tau)
= C\times \frac{\Sigma_{\rm GMC}(\overline{R}_{\rm GC},\tau)}{\widehat{J_z}(\overline{R}_{\rm GC},\tau)\Sigma_* (\overline{R}_{\rm GC},\tau)},
\end{equation}
which can be solved by straightforward numerical integration, if $\Sigma_{\rm GMC}(\overline{R}_{\rm GC},\tau)$ and $\Sigma_* (\overline{R}_{\rm GC},\tau)$ are known. Note that if $\Sigma_{\rm SFR}$ and $\Sigma_*$ are constant with time, this recovers the expected $J_z \sim t^{0.5}$ relation in the case of a static Milky Way disk. 

To estimate how the density of scatterers, $\Sigma_{\rm GMC}$ evolves with time and epoch, we assume that  $\Sigma_{\rm GMC}$ scales with star formation-rate density $\Sigma_{\rm SFR}  (\overline{R_{\rm GC}}, \tau)$ via the Schmidt-Kennicutt law with $\Sigma_{\rm GMC} \sim \Sigma_{\rm SFR}^{1/\alpha_{\rm KS}}$ and adopt $\alpha_{\rm KS} = 1.0$, the measured Schmidt-Kennicutt index for the cold molecular clouds \citep{big08,ken12}. We adopt the global star-formation history of the Galactic disk from \citet{fra18}, where we consider both an exponential decay of the SFR, as well as an inside-out growth of the disk. More precisely, we assume
\begin{eqnarray}
\Sigma_{\rm SFR} (\overline{R_{\rm GC}}, t_{\rm look-back})  =  \exp{\bigl ( -\frac{(12-t_{\rm look-back}}{\tau_{\rm SFR}} \bigr )} \nonumber \\ 
 \times \frac{1}{R_{\rm exp}(t_{\rm look-back})} \exp\Big(- \frac{\overline{R}_{\rm GC}}{R_{\rm exp}(t_{\rm look-back})}\Big),
\end{eqnarray}

\noindent
and adopt the Milky Way global parameters $\alpha_{R} = 0.4$, $R_{{\rm SFR},0} = 3\,{\rm kpc}$, $\tau_{\rm SFR} = 6.5\,{\rm Gyr}$, and
\begin{equation}
\label{eq:scale-length-evolution}
R_{\rm exp} (t_{\rm look-back}) = R_{{\rm SFR},0} \, \Big( 1- \alpha_{R} \cdot \frac{t_{\rm look-back}}{8\,{\rm Gyr}} \Big),
\end{equation}

The scaling $C$ sets the overall scale for $\widehat{J_z}$, but will not change the power-law index in the resulting $\widehat{J_z}(\tau)\propto\tau^\gamma$. We find that $C=15$ provides a good agreement between the data and the predictions for all radii. For the stellar mass evolution, we assume an exponential profile with a scale length inside-out evolution similar to Eq.~(\ref{eq:scale-length-evolution}) with the same $\alpha_R$, but we adopt the current stellar mass scale length to be $R_{{\rm *},0} = 2\,{\rm kpc}$ \citep[e.g.,][]{bov13} in lieu of $R_{{\rm SFR},0} = 3\,{\rm kpc}$.

Fig.~\ref{fig7} shows the numerical evaluation of the differential equation in Eq.~(\ref{eq:heating_equation}) at different radii, assuming the initial condition $\widehat{J_z}(t=0) = 1\;$kpc$\,$km/s, as per Section~\ref{sec:results}. Individual lines in color show the $J_z(t)$-evolution of various stellar populations, color-coded by the birth time. We assume the smallest birth time to be $1\,$Gyr, mimicking some of the youngest stars in the red clump sample. The figure shows that populations born earlier experienced more heating in $\widehat{J_z}$ than more recently born populations, reflecting the higher $\Sigma_{\rm GMC}$ and lower $\Sigma_*$ at the time. Each evolutionary track ends at the present epoch with $\widehat{J_z}(t=\tau)$. Remarkably with just scaling $C$ the predictions, the ensemble of endpoints of the evolution tracks agree well with the observed $\widehat{J_z}(\overline{R}_{\rm GC},\tau)$.

For a more quantitative comparison we proceed to fit the endpoints of these tracks with Eq.~(\ref{eq-model-start}), as proxy for the observable mean $\widehat{J_z}(\tau)$, after perturbing the final $\widehat{J_z}$ of each track by $\sigma_{Jz}/J_z = 15\%$.  The black lines illustrate the best-fitted power-law for the analytic prediction, with the power law index is shown in the top left corner of each panel. The red dashed lines are the mean of the best global fit from the red clump sample presented in Section~\ref{sec:results}; note that this figure differs slightly from Fig.~\ref{fig5} because, unlike  Fig.~\ref{fig5}, we have not folded in the selection function here. 

The best-fit power-law from this physically inspired model has a range of $\gamma \simeq 1$, remarkably resembling the observed $\widehat{J_z}(\tau)$, even though all individual populations are evolved with $J_z \sim t^{0.5}$. This analytic model also shows an increase in $\gamma$ with radius. When the inside-out growth is included, the disk stars at the outer disk start with a small restoring force due to the small radial scale length of the stellar disk. As the disk ages and grows, the increment in the restoring force at the outer disk is more drastic than the inner disk. As a result, as shown in the right panel of Fig~\ref{fig7}, there are larger separations between individual evolutionary tracks, and the slower evolution in $J_z$ for the younger populations means that the observed $J_z \sim \tau$ relation will favor a ``long''-term evolutionary behavior where the $J_z$ increases more drastically for an older look-back time. And this causes the observed $\gamma$ appeared to be larger at the outer disk.

It is encouraging how well this simple model does without fine-tuning: perhaps gradual orbit scattering is indeed the dominant source driving the $J_z$ evolution across the Milky Way disk, at least for $R \lesssim 14$~kpc. But we note that this models encompasses the scenario where in-plane heating (or blurring) can happen through a range of processes, such as spiral arms and bars perturbations, with GMCs simply isotropizing the in-plane non-circular motion of stars. 

This analysis may also imply that the influence of satellite bombardment, at least for $3\;{\rm kpc} < \overline{R}_{\rm GC} < 14\,$kpc need not be a dominant source of vertical heating. If there is a major merger impacted the Milky Way thin disk, one would expect the heating rate for that specific population should be higher than usual, which would then imprint as a higher $\gamma$ observed for the same reasoning that we laid out above. But if the Milky Way thin disk only went through a steady stream of minor mergers, this might not manifest itself in the observed $\gamma$, but we will defer a more careful comparison with detailed simulations to future studies.

%
%
%
%
%
%

\subsection{The present-day vertical actions as birth properties?}

While our vertical heating model provides an excellent explanation of the data, we assume that (low-$\alpha$) stars were born at a constant birth-temperature.  This assumption is supported by some simulations \citep[e.g.,][]{mar14}, but contested by others \citep[e.g.,][]{bro04,bro06,bir13}. We could not rule out the possibility that stars could merely be born kinematically hotter in the past, and the evolution of $J_z$ is simply a consequence of the cooling of the ISM. The latter is known to play a significant role for the older thick disk when the ISM gas of the Milky Way is still settling down \citep{bou09,bir13,gra16}, what is known as the ``upside-down'' formation.  

Nonetheless qualitatively, external galactic observations have shown that the velocity dispersion, $\sigma_z$ at any galactic annulus scales as $\sigma_z^2 \Sigma_{\rm gas} \sim \Sigma_{\rm SFR}$ \citep[e.g., fig4 in][]{tam09}, and thus, if the Galactic vertical structure for the thin disk is dominated by the upside-down formation, we would expect, $J_z \sim \sigma_z^2 \sim \Sigma_{\rm SFR}/\Sigma_{\rm gas}$. To this end, we have tested that, let $\Sigma_{\rm SFR}$ to be a exponential profile with an exponential time-scale $\tau_{\rm SFR} = 3-8\,$Gyr, both assuming $\Sigma_{\rm gas}$ to be constant, i.e., $J_z \sim \exp(-t_{\rm look-back}/\tau_{\rm SFR})$, and assuming $\Sigma_{\rm gas} \sim \Sigma_{\rm SFR}^{1/\alpha_{\rm KS}}$, i.e.  $J_z \sim \exp(-[(1-1/\alpha_{\rm KS}) \cdot t_{\rm look-back}/\tau_{\rm SFR}]$, provides a far poorer fit to the observed red clump $J_z-\tau$ trend than the fitted trend in Fig.~\ref{fig7}. Fundamentally, an exponential profile is just not a good approximation for $J_z \sim \tau^{\sim 1}$ which leads us to conclude that the evolution of the vertical structure as measured in this study is better explained by GMC scattering.

\vspace{1cm}

%
%
%
%
%
%

\section{Conclusion}
\label{sec:conclusion}

We have quantified the global vertical temperature of the Milky Way's stellar disk as a function of stellar age, $J_z (\Rbar, \tau)$, spanning the Galactocentric radii from 3$\,$kpc to 14$\,$kpc and ages $\tau < 8$~Gyr.  To this end, we combined a pristine subset of APOGEE red clump stars previously derived in \citet{tin18a} and proper motions from the recent Gaia DR2 \citep{gai18}, yielding a sample of $\sim 20,000$ stars with the precision of $7\%$ in distance, $25\%$ in stellar age and $15\%$ vertical action across the Milky Way. We chose to model the vertical action, $J_z(\Rbar,\tau)$ instead of the classical age-velocity dispersion relation, as $J_z$'s adiabatic invariance makes the results more interpretable. Our finding can be summarized as followed:
\begin{itemize}
\item The full distribution of $J_z$ at any given $(\Rbar,\tau)$, can be very well approximated by an isothermal distribution $p(J_z) \sim \exp(-J_z/\widehat{J_z})$. 

\item The best-fitted $\widehat{J_z} (\overline{R}_{\rm GC},\tau)$ then informs about the evolution of $J_z$ at different radius and time. But, importantly, $\widehat{J_z} (\overline{R}_{\rm GC},\tau)$ does not reflect the evolutionary paths of any stellar population {\it per se}. We parametrize $\widehat{J_z}(\overline{R}_{\rm GC},\tau) \equiv \widehat{J_{z,0}} (\overline{R}_{\rm GC})\  +\  \DJz(\overline{R}_{\rm GC})\cdot (\tau/1\,{\rm Gyr})^{\gammRC}$.

\item In fitting this parameterized model to the data, we account for the selection function in $J_z$, $S(J_z)$, well approximated by a broken power-law. We also find that accounting for the observational uncertainties in this fitting matters.

\item  We find the birth actions $\widehat{J_{z,0}}$ to be consistent with the expectation from the ISM or star cluster velocity dispersion, for $\overline{R}_{\rm GC} < 10\,$kpc: $\widehat{J_{z,0}} = 1\,$kpc km/s. At larger radii, the birth action increases to $\widehat{J_{z,0}} = 8\,$kpc km/s at $R_{\rm GC} = 14\,$kpc, which might indicate that stars were formed under Galactic warp or flare beyond $10\,$kpc, enabled by the lower disk self-gravity at these radii.

\item The increase of $J_z$ with stellar age for the last Gyr, traditionally interpreted as the ``heating rate'' $\DJz$, shows a constant value $\sim 1.6\;{\rm kpc}\;{\rm km/s}\;{\rm Gyr}^{-1}$, with only mild dependence with radius. But its power law scaling with age increases with Galactocentric radius, from $\gamma \simeq 1$ at $\overline{R}_{\rm GC} = 3\,$kpc to $\gamma \simeq 1.3$ at $\overline{R}_{\rm GC} = 14\,$kpc.

\item Our constraint on $\gamma$ in the Solar neighborhood is consistent with the GCS \citep{hol09} value $\sigma \sim \tau^{0.53}$ (or, $\gamma \sim 1.06$).

\item To cast the global empirical constraint on $\widehat{J_z} (\overline{R}_{\rm GC},\tau)$ we derived, in terms of actual heating scenarios, we present a simple analytic model: we assume that the vertical heating of all stellar populations is dominated by orbit scattering (e.g., from GMCs), with $J_z\propto t^{0.5}$. But we account for changes in the scattering amplitude with epoch,  due to an exponential decrease of SFR in the Milky Way and an inside-out growth of the Milky Way stellar disk. We show that such a model can reproduce the range of power-law indices $\gamma$ in observed present-day age-action relation $\widehat{J_z} (\tau)$, despite $\gamma > 0.5$ at all radii. The result suggests that orbit scattering (from GMCs) might be the dominant source of disk vertical heating for much of the low-alpha disk over the last 8$\,$Gyrs. The thin disk is unlikely to have gone through any dramatic major merger, although we could not rule out indirect influence from the spiral arms, the Galactic bar and the impact from minor mergers.
\end{itemize}

While the unique data from Gaia and APOGEE allows for the measurement of the global heating rate in the form of vertical action across the Milky Way disk, this study undoubtedly only explored a small part of what could potentially be attained by combining Gaia with large spectroscopic surveys like APOGEE, Galah, Gaia-ESO, and LAMOST. The full potential can only be realized if we incorporate the vertical heating model into a complete description of the Milky Way where more aspects of the Milky Way are studied and constrained simultaneously. And with only that, we might finally do justice to the data, and a fuller understanding of the formation of the Milky Way might eventually come to focus.

%
%
%
%
%

\section{Acknowledgments}
\label{sec:acknoledgements}

YST is supported by the NASA Hubble Fellowship grant HST-HF2-51425.001 awarded by the Space Telescope Science Institute, the Carnegie-Princeton Fellowship and the Martin A. and Helen Chooljian Membership from the Institute for Advanced Study in Princeton. HWR's research contribution is supported by the European Research Council under the European Union's Seventh Framework Programme (FP 7) ERC Grant Agreement n. [321035] and by the DFG's SFB-881 (A3) Program. This project was developed in part at the 2018 NYC Gaia Sprint, hosted by the Center for Computational Astrophysics of the Flatiron Institute in New York City. This work has made use of data from the European Space Agency (ESA) mission {\it Gaia} (\url{https://www.cosmos.esa.int/gaia}), processed by the Gaia Data Processing and Analysis Consortium (DPAC, \url{https://www.cosmos.esa.int/web/gaia/dpac/consortium}). Funding for the DPAC has been provided by national institutions, in particular the institutions participating in the Gaia Multilateral Agreement. The Sloan Digital Sky Survey IV is funded by the Alfred P. Sloan Foundation, the U.S. Department of Energy Office of Science, and the Participating Institutions and acknowledges support and resources from the Center for High- Performance Computing at the University of Utah. This publication makes use of data products from the Wide-field Infrared Survey Explorer, which is a joint project of the University of California, Los Angeles, and the Jet Propulsion Laboratory/California Institute of Technology, funded by the National Aeronautics and Space Administration.

%
%
%
%
%
%

\appendix
\section{\\Selection function in $J_z$}

In this appendix, we describe the details of the $J_z$ selection function, its analytic approximation, and its use in modifying the data likelihood.

To make the selection function in $J_z$ more tractable, we assume the APOGEE sample at $|z_{\rm GC}| < 0.5\,$kpc is unbiased with respect to $z$, i.e., we assume that APOGEE uniformly samples stars from $|z_{\rm GC}| < 0.5\,$kpc. We argue that this approximation is reasonable as the APOGEE densely target the Milky Way midplane. But we emphasize that this is strictly an approximation for numerical convenience. A more proper account of the selection function involves $S(l,b)$, taking into account the APOGEE pointing. We will defer such detailed modeling to future studies.

The selection function in $J_z$ should depend on the radius because the Galactic potential and hence the vertical restoring force felt by stars are different at different Galactocentric radii, but we assume that the selection function is independent of time due to the short vertical oscillation timescale. Therefore, even young stars should be mostly phase-mixed in the vertical direction. With these in mind, to evaluate $S(J_z|\overline{R}_{\rm GC})$, it suffices to estimate simply the fraction of time for a star that has a vertical action of $J_z$ and at radius $\overline{R}_{\rm GC}$ to spend in the observable volume of $|z_{\rm GC}| < 0.5\,$kpc. As the vertical action fully characterizes the vertical orbit oscillation, the selection function is insensitive to radial action $J_R$ of the stars.

We adopt the Milky Way potential from {\sc galpy} and assume $R_\odot = 8.2\,$kpc, $v_{\circ}(R_\odot) = 240\,$km/s as before. At each radius, we calculate the corresponding circular velocity $v_{\rm circ}(\overline{R}_{\rm GC})$ from the potential and orbit integrate a dense grid of stars with circular orbits, starting in the midplane $z_{\rm GC}=0$, but with different vertical velocity $v_z$. In other words, we orbit integrate stars with initial velocities $(v_R, v_\phi, v_z) = (0, v_{\rm circ}(\overline{R}_{\rm GC}), v_z)$ for a large range of $v_z$. These stars probe, at each Galactocentric radius, a wide range of vertical motions, which vertical action $J_z$ and orbits will then inform how much time stars with different $J_z$ spend in the midplane. The $J_z$ of these stars are calculated as described in Section~\ref{sec:data}, and we evaluate $S(J_z|\overline{R}_{\rm GC})$ to be the fraction of the total integration time such that the orbits are within $|z_{\rm GC}| < 0.5\,$kpc. The left panel of Fig.~\ref{fig3} shows the estimated $S(J_z|\overline{R}_{\rm GC})$ with this ``numerical'' method.

The left panel shows that the selection function varies for different Galactocentric radii. This is not surprising because, in the inner disk, stars are subjected to more vertical restoring force. As a result, more stars with higher $J_z$ spend their entire time (i.e., $S(J_z|\overline{R}_{\rm GC} = 100\%$) at $|z_{\rm GC}| < 0.5\,$kpc. But even at $R_{\rm GC} = 2\,{\rm kpc}$, only stars with vertical actions smaller than $J_z < 25\,{\rm kpc}\,{\rm km/s}$ are complete. As shown in the panel, the selection function drops precipitously after that, reaching as low as $S(J_z|\overline{R}_{\rm GC}) = 10\% - 20\%$ for $J_z = 100\,{\rm kpc}\,{\rm km/s}$, the largest $J_z$ we observe in the data. The selection function is even more drastic at the outer radii, and only stars with the smallest $J_z$ ($J_z \lesssim 5\,{\rm kpc}\,{\rm km/s}$) are complete. It is therefore important to stress that, due to the high incompleteness of the observation, especially a strong dependence with $J_z$ and $\overline{R}_{\rm GC}$, a robust vertical heating history can only be inferred with a proper selection function included. In particular, instead of comparing the data to $p(J_z|\overline{R}_{\rm GC},\tau,{\bf p})$ predicted by the isothermal distribution function, where ${\bf p}$ are the model parameters, we have the ``true'' model prediction to be
\begin{equation}
\widetilde{p}(J_z|\overline{R}_{\rm GC},\tau,{\bf p}) = \frac{1}{C(\overline{R}_{\rm GC},\tau,{\bf{p}})} p(J_z|\overline{R}_{\rm GC},\tau,{\bf{p}}) \cdot S(J_z | \overline{R}_{\rm GC})
\end{equation}

\noindent
where $C(\overline{R}_{\rm GC},\tau,{\bf{p}})$ is the normalization of the predicted distribution function,
\begin{equation}
C(\overline{R}_{\rm GC},\tau,{\bf{p}}) = \int_0^{J_{z,{\rm max}}} {\rm d}J_z \; p(J_z|\overline{R}_{\rm GC},\tau,{\bf{p}}) \cdot S(J_z | \overline{R}_{\rm GC}).
\end{equation}

But evaluating such normalization through a numerical grid of $S(J_z | \overline{R}_{\rm GC})$ could be computationally prohibitive since we need to evaluate the normalization for all stars (with different $\overline{R}_{{\rm GC},i}$ and $\tau_i$) at each step in the MCMC likelihood call. Therefore, a sensible approach is to approximate $S(J_z | \overline{R}_{\rm GC})$ with an analytic expression with which $C(\overline{R}_{\rm GC},\tau,{\bf{p}})$ can be evaluated analytically. In particular, we find that $S(J_z|\overline{R}_{\rm GC},\tau)$ can be well approximated by

\begin{equation}
S(J_z|\overline{R}_{\rm GC}) =
   \left\{
   \begin{aligned}
     1, &\qquad {\rm for\;} J_z \leq \alpha(\overline{R}_{\rm GC}) \\
     \beta(\overline{R}_{\rm GC}) \cdot J_z^{-\xi(\overline{R}_{\rm GC})}, &\qquad {\rm for\;} J_z > \alpha(\overline{R}_{\rm GC})
   \end{aligned}
   \right.
\end{equation}

\noindent
\begin{eqnarray}
\alpha(\overline{R}_{\rm GC}) &=& -0.0132 \overline{R}_{\rm GC}^3 + 0.5033 \overline{R}_{\rm GC}^2 -6.7793 \overline{R}_{\rm GC} + 36.4815, \nonumber \\
\beta(\overline{R}_{\rm GC}) &=& 0.0013 \overline{R}_{\rm GC}^4 - 0.0609 \overline{R}_{\rm GC}^3 + 1.0425 \overline{R}_{\rm GC}^2  -8.1934 \overline{R}_{\rm GC} + 27.7762, \nonumber \\
\xi(\overline{R}_{\rm GC}) &=& -0.000058 \overline{R}_{\rm GC}^3 + 0.0030 \overline{R}_{\rm GC}^2 - 0.0563 \overline{R}_{\rm GC} +1.0056.
\end{eqnarray}

The middle panel of Fig.~\ref{fig3} illustrates the selection function evaluated with this analytic formula, and the right panel shows the difference between the numerical selection function as plotted in the left panel and this analytic approximation. As shown, the two agree very well, except at the transition region where we begin to have $S(J_z|\overline{R}_{\rm GC}) < 1$. But this error only affects a small fraction of the observation data, and we checked that this does not modify much the inference of this study. But the analytic approximation allows the evaluation of the normalization much faster without which we would not be able to perform a Monte Carlo integration of the observable uncertainties. In particular, with this analytic approximation, one can show that the distribution function normalization reads

\begin{eqnarray}
C(\overline{R}_{\rm GC},\tau,{\bf{p}}) &=& \widehat{J_z}(\overline{R},_{\rm GC}\tau,{\bf{p}}) \cdot \Bigg[ 1- \exp \bigg( -\frac{\alpha (\overline{R}_{\rm GC})}{\widehat{J_z}(\overline{R}_{\rm GC},\tau,{\bf{p}})} \bigg) -  \beta(\overline{R}_{\rm GC}) \cdot J_{z,{\rm max}}^{-\xi(\overline{R}_{\rm GC})}  \cdot \bigg( \frac{J_{z,{\rm max}}}{\widehat{J_z}(\overline{R}_{\rm GC},\tau,{\bf{p}})}  \bigg)^{\xi(\overline{R}_{\rm GC})} \times \Gamma\bigg(1-\xi(\overline{R}_{\rm GC}),  \frac{J_{z,{\rm max}}}{\widehat{J_z}(\overline{R}_{\rm GC},\tau,{\bf{p}})} \bigg) \nonumber \\
&&\qquad\qquad\qquad\qquad +  \beta(\overline{R}_{\rm GC}) \cdot \alpha(\overline{R}_{\rm GC})^{-\xi(\overline{R}_{\rm GC})} \cdot  \bigg( \frac{\alpha(\overline{R}_{\rm GC})}{\widehat{J_z}(\overline{R}_{\rm GC},\tau,{\bf{p}})}  \bigg)^{\xi(\overline{R}_{\rm GC})}  \times \Gamma\bigg(1-\xi(\overline{R}_{\rm GC}),  \frac{ \alpha(\overline{R}_{\rm GC})}{\widehat{J_z}(\overline{R}_{\rm GC},\tau,{\bf{p}})} \bigg) \Bigg]
\end{eqnarray}

\noindent 
where $\Gamma$ is the upper incomplete gamma function,
\begin{equation}
\Gamma(a,x) = \int_x^\infty {\rm d}t \; t^{a-1} e^{-t},
\end{equation}

\noindent
and here we choose the upper limit of the integral $J_{z,{\rm max}} = 1000\;{\rm kpc}\;{\rm km/s}$. Finally, given the observation data $\{J_z, \overline{R}_{\rm GC}, \tau\}_i$ of the red clump sample, the model parameters ${\bf p}$ can be inferred through Bayesian inference via MCMC as described in the main text.

%
%
%
%
%
%

\end{CJK*}

\vspace{1cm}
\bibliography{biblio.bib}

\end{document}